\documentclass[aps,prl,twocolumn,superscriptaddress,showpacs]{revtex4-1}
\usepackage{graphicx}
\usepackage{array} 
\usepackage{hyperref} 
\usepackage{amsmath,amssymb} 
\usepackage{amsfonts}
\usepackage{bm}
\usepackage{float} 

\begin{document}
\title{Turbulence induces clustering and segregation of non-motile, 
buoyancy-regulating phytoplankton} \date{\today}

\author{Matteo Borgnino} \affiliation{Department of Physics and INFN,
Universit\`a di Torino, via P. Giuria 1, 10125 Torino, Italy} 
\author{Jorge Arrieta} \affiliation{Instituto Mediterr\'aneo de Estudios Avanzados, IMEDEA, UIB-CSIC, Esporles, 07190, Spain} 
\author{Guido Boffetta} 
\author{Filippo De Lillo} 
\affiliation{Department of Physics and INFN, Universit\`a di Torino, via P.  Giuria 1, 10125 Torino, Italy} 
\author{Idan Tuval} \affiliation{Instituto Mediterr\'aneo de Estudios Avanzados, IMEDEA, UIB-CSIC, Esporles, 07190, Spain}

\begin{abstract} 
Turbulence plays a major role in shaping marine community structure as it
affects organism dispersal and guides fundamental ecological interactions. Below
oceanographic mesoscale dynamics, turbulence also impinges on subtle
physical-biological coupling at the single cell level, setting a sea of chemical
gradients and determining microbial interactions with profound effects on scales
much larger than the organisms themselves. It has been only recently that we
have started to disentangles details of this coupling for swimming
microorganisms. However, for non-motile species, which comprise some of the most
abundant phytoplankton groups on Earth, a similar level of mechanistic
understanding is still missing. Here we explore by means of extensive numerical
simulations the interplay between buoyancy regulation in non-motile
phytoplankton and cellular responses to turbulent mechanical cues. 
Using a minimal mechano-response model we show how such mechanism would
contribute to spatial heterogeneity and affect vertical fluxes and trigger community segregation.
\end{abstract}


\maketitle
	
\section{Introduction} \label{sec:intro}
	
The spatial distribution of aquatic microorganisms has profound effects on the
ecology of our oceans \cite{Williams2011,Mitchell2008} affecting fundamental
ecological interactions, population stability, species
diversity~\cite{Legendre1989} and, hence, affecting the functioning of whole
marine food webs~\cite{Polis1997}. Highly sparse non-uniform spatial
distributions, or patchiness, has distinct origins at different scales: while at
the mesoscale it is mostly driven by reproduction, grazing, nutrient
availability \cite{Mackas1985} and advection by currents~\cite{Martin2003}, at
smaller scales (from the scale of the cell up to the order of the meter)
 the interplay between biological and physical factors plays a
major role. These include microorganismal motility and its interaction with
fluid flows and, among other processes, it shapes encounter
rates~\cite{kiorboe2008,Visser2006}, the formation of thin
layers~\cite{Durham2012}, cell clustering~\cite{Durham2013}, and segregation~\cite{Margalef1978}.
Small scale patches also serve as hotspots of microbial activity facilitating, for instance, 
interaction with bacteria in the "phycosphere",  influencing global carbon and nutrient cycling, 
and regulating ecosystem productivity~\cite{Azam2007,Seymour2017}.
	
Recent efforts have clearly established that motile marine microorganisms are
patchily distributed in the presence of turbulent flows \cite{durham2013turbu}.
Cell motility and their response to chemical and mechanical landscapes conspire
with fluid flows to accumulate and disperse cells in different spatial
environments. However, and despite non-motile species comprising two of the most
important ecological groups in the ocean (cyanobacteria, essential for nitrogen
fixation \cite{Berman-Frank2003,Zehr2010}, and diatoms, carrying out about one fifth of the total photosynthesis
on Earth~\cite{Malviya2016}), a largely unanswered question concerns the responses of non-motile
cells to the same turbulent cues and how it affects their sinking dynamics, of
paramount importance for global biogeochemical cycles.

Motivated by the physiological regulation of buoyancy prevalent in non-motile
phytoplankton species \cite{Gross1948, Waite1992, Walsby94, arrieta2015microscale}, here
we investigate, by means of direct numerical simulations, the dynamics of active
but non-motile cells in a three-dimensional turbulent flow. In particular, we
focus on cells response to mechanical stresses as those locally induced by fluid
forces. Although non-motile species possess the required mechanosensitive machinery to display 
rapid active responses to imposed mechanical stresses (triggering, for instance, the production 
of cytosolic Ca$^{2+}$~\cite{Falciatore2000}), the effect of hydrodynamic stresses in buoyancy 
regulation has been largely overlooked. Overcoming adverse environmental conditions, including light 
and nutrient limitations, has been considered as the most relevant driver for buoyancy 
regulation~\cite{Falciatore2002, Erga2009, Gemmell2016}. 
Whilst certainly important, more recent work suggests this is only part of the story: 
even under nutrient replete conditions, transcriptional analysis reveals rapid changes in gene expression 
solely associated to the exposure to turbulent flows. These include an increase in the fatty acid (FA) 
biosynthesis pathways (which may also serve as buoyancy regulators) and other determinants of 
cellular metabolic state~\cite{Amato2017}. Moreover, fast and active physiological responses 
are also known to directly regulate cells instantaneous sinking speeds~\cite{Gemmell2016}. 
While none of these studies disentangle the mechano-transduction pathway linking physiological 
responses to changes in cell density, they unambiguously show that non-motile cells 
are able to perceive, and actively respond, to mechanical stimuli in short times. 
Since buoyancy regulation is the only known mechanism for non-motile species to control their 
position in the water column, the above results render mechano-induced buoyancy regulation 
as a plausible hypothesis that serves as the starting point for our minimal model.

Here we show that, in contrast to passive tracers, a simple law for buoyancy regulation leads to cell
clustering and species segregation, and we demonstrate how these processes
depend on physical parameters such as the cells' settling speed. Finally, we
discuss its implication for the ecology of marine phytoplankton.
	
\section{Mathematical model} 
\label{sec:model} 
We consider the motion of small
spherical particles of radius $a$ and variable density $\rho_p$ immersed in
three-dimensional turbulent flows described by the incompressible Navier-Stokes
equations 
\begin{equation}
	\begin{aligned} \label{eq:NS} 
	\nabla \cdot {\bm u} &= 0\\ 
	\partial_t {\bm u} + {\bm u} \cdot {\bm \nabla} {\bm u} &= 
	-{1 \over \rho} {\bm \nabla} p + \nu \nabla^2 {\bm u} + {\bm f}
	&&
	\end{aligned}
\end{equation}
Here $\bm{u}(\bm{x},t)$ is the fluid velocity, $p(\bm{x},t)$ the 
pressure, $\rho$ the uniform fluid density and
$\nu$ its kinematic viscosity. The forcing term, ${\bm f}$, represents a
zero-mean, temporally uncorrelated, Gaussian forcing which injects energy at
large scales at a given rate, $\varepsilon$, necessary to sustain a
statistically stationary state. Together with the kinematic viscosity, the
energy injection rate defines the Kolmogorov scales for the length,
$\eta=(\nu^3/\varepsilon)^{1/4}$, time, $\tau_{\eta}=(\nu/\varepsilon)^{1/2}$,
and velocity, $u_{\eta}=\eta_k/\tau_k=(\nu \varepsilon)^{1/4}$
\cite{frisch1995turbulence}. These scales will be used to make physical 
quantities dimensionless.

Small spherical particles follow the Maxey-Riley equation~\cite{Maxey1983}. 
In our case, as the Reynolds number $Re_p$ based on the particles radius $a$ and the
characteristic velocity $U_a$ (i.e. the maximum between particle sedimentation velocity
and the characteristic velocity fluctuation at the scale of the particle),
is very small, $Re_p=U_a a/\nu\ll 1$, derivatives computed following the particles, $d/dt$,
are well approximated by derivatives along fluid streamlines, $D/D t$.
Furthermore, we neglect Faxen corrections and the Basset history terms, 
following standard approaches justified for very small relaxation time 
\cite{Provenzale1999,Babiano2000,Daitche_2014}. 
Finally, the acceleration of the particle is written as
\begin{equation} 
{d {\bm u}_p \over dt} = \beta \frac{d \bm{u}}{d t} - 
\frac{\bm{u}_p-\bm{u}}{\tau_p}-(1 - \beta)g \hat{\bm k},
\label{eq:2.1} 
\end{equation}
where ${\bm u}_p=d{\bm x}_p/dt$ is the particle velocity,
$\beta=3\rho/(2\rho_p+\rho)$ is ratio of the fluid density to the particle
density $\rho_p$, $\tau_p=a^2/(3\nu\beta)$ is the Stokes relaxation time
and $g$ represents the acceleration of gravity. 

We further simplify the equations taking into account the fact that particles
are almost neutrally buoyant ($\beta \simeq 1$), that 
$g \gg |d{\bm u}/dt|$ - a typical condition in the ocean - and that 
the Stokes time (order $10^{-3} s$ for \textit{Stephanodiscus rotula}
$a\sim 50\cdot10^{−6} m$, $\rho \sim 1020 kg/m^3$ \cite{Reynolds2006})
is usually much smaller that the smallest time scale in the flow 
(range from $0.1 s$ in coastal regions $\varepsilon \sim 10^{-4} W/kg$ 
to $10 s$ in the open ocean $\varepsilon \sim 10^{-8} W/kg$
 \cite{Thorpe2007}).
Under these conditions the equation of motion
(\ref{eq:2.1}) reduces to an equation for particle position
\begin{equation}
{d {\bm x}_p \over dt} = {\bm u}_p =
{\bm u} - v_s \hat{\bm k} 
\label{eq:2.2} 
\end{equation}
where $v_s=(1-\beta)\tau_p g$ is the particle sinking speed in still fluid
and $\hat{\bm k}$ represents the vertical direction. 
	
For particles with constant density, $\rho_p$, the motion described by
(\ref{eq:2.2}) is identical to that of ideal fluid tracers in the presence of an
additional constant vertical drift due to buoyancy forces. 
Dynamics under these conditions cannot
produce particle clustering as the relative motion between particles is
identical to that of fluid elements. More formally, in this case the effective
velocity field ${\bm u}_p$ is divergence free and therefore the rate of
contraction in physical space is zero~\cite{Ott2002}. 
This dynamics does not hold in the
case of particles which are able to regulate their density (buoyancy) as is the
case of diatoms and cyanobacteria. In the following we will describe our model
for buoyancy regulation in response to fluid mechanical stresses, following that
discussed in \cite{arrieta2015microscale}. To be specific, we assume that the
particle density is dependent on the norm $S$ of the local strain tensor 
${\cal S}_{ij}=1/2(\partial_{i}u_{j}+\partial_{j}u_{i})$, and
in particular, we employ the
Frobenius norm $S=[tr({\cal S}{\cal S}^{t})]^{1/2}$. Since a change in particles
density corresponds to a change in sedimentation velocity, the latter will
depend on the flow strain rate computed on the particle position
$v_{s}=v_{s}(S_{p})$. As the detailed mechanism responsible for how
intracellular responses translates into the regulation of buoyancy is still
unclear, we will analyse two possible scenarios differing in the sign of the
response: we refer to cells whose density decreases (increases) with the
mechanical stresses as \textit{shear-thinning} (\textit{shear-thickening})
respectively \cite{arrieta2015microscale}.
	
To model how cells sedimentation velocity changes with the local strain rate we
take inspiration from Michaelis-Menten kinetics. We choose a response function
of the from $f(S)=S/(S+S_{H})$, where $S_{H}$ is the strain rate half
saturation constant. Examples of this response functions are shown in
Figure~\ref{fig:S_eul} together with a typical steady distribution of strain rate
in a turbulent flow obtained from the integration of \eqref{eq:NS}. We further
assume that the density of the particle varies linearly with the response
function $f(S)$ in the range $\rho \leq \rho_{p} \leq 2\rho$
\cite{arrieta2015microscale}, where the minimum density corresponds to a
neutrally buoyant particle. Thus, the density law for the
shear-thinning case is $ \rho_p=\rho [2-f(S)]$, whereas for the
shear-thickening case is $\rho_p=\rho[1+f(S)]$. These density laws provide
the variation of the still fluid sedimentation velocity with the norm of the
strain rate for the shear-thinning case,
$v_{s}(S)=\left[1-\frac{S}{S+S_{H}}\right]v_{s,max}$ and the
shear-thickening case $v_{s}(S)=\frac{S}{S +S_{H}}v_{s,max} $, where
$v_{s,max}=2a^2g/(9\nu)$ is the still fluid sedimentation velocity 
at the maximum density $\rho_p=2 \rho$.
Note that at the half-saturation constant we have
$v_s(S_H)=v_{s,max}/2$. 
The choice $2 \rho$ for the maximum density is not restrictive
since $v_s$ is given by a combination of $\rho$ and $a$ and the results
for a different maximum density would be equivalent to those obtained 
for a cell of different size.
We emphasize that in deriving the model of buoyancy we
have assumed that the particle regulates its buoyancy immediately. This is in
agreement with the characteristic time scale measured in the back-and-forth
transition in the sinking rate in \cite{Gemmell2016} and the measured time
response to other environmental signals \cite{Falciatore2000}. 
It is important to clarify that the buoyancy model derived therein does not include
any adaptive response to mechanical stresses and that, in \cite{Gemmell2016}, 
buoyancy regulation is seen as a mechanism able to enhance nutrient uptake 
by altering the nutrient-deplete boundary layer around the cell, regardless of the external flow.
However, this does not mean that this process is not relevant or  
does not take place within a turbulent environment 
(where the boundary layer argument at the basis of \cite{Gemmell2016} would be less stringent).
Indeed, it was recently shown that
motile phytoplankton are able to actively modify their migration strategy to
evade turbulent layers~\cite{Sengupta2017} while non-motile phytoplankton (e.g.
diatoms) modify their gene expression to trigger energy storage pathways when
exposed to turbulent flows~\cite{Amato2017}.
	
\section{Numerical results} 
\label{sec:results} 
We have performed a numerical investigation of the statistical properties
of several populations of both shear-thinning and shear-thickening cells. 
By means of direct numerical simulations of the NS equations \eqref{eq:NS} 
using a fully-dealiased pseudo-spectral code \cite{Boyd2001}
we obtain the incompressible velocity field.
Statistical stationarity of the flow is guaranteed by a
white-in-time forcing ${\bm f}$ acting at large scale only.
Simulations are done at three different resolutions $N=64, 128, 512$ 
($N$ is number of grid points 
per side on the periodic cube of size $L_B$) corresponding to three 
different Reynolds numbers or turbulence intensities. Resolutions 
are chosen such that the maximum wavenumber available $k_{\rm
max}$ satisfies the relation $k_{\rm max} \eta> 1.8$ to guarantee sufficient
accuracy at small scales at the different turbulence intensities.
In stationary conditions, a population of $N_c$ cells is
initialized with uniform random positions ${\bm x_{p}}$ in the domain. 
Particle trajectories are obtained by the simultaneous integration of 
\eqref{eq:NS} and \eqref{eq:2.2} where the velocity field
and the strain rate at the cell positions are obtained by a third-order
polynomial interpolation. After the particles distribution has reached a
statistically steady state, we collect data for several large-scale eddy
turnover times to ensure statistical convergence.

\begin{figure}[h!] 
\centering 
\includegraphics[scale=0.7]{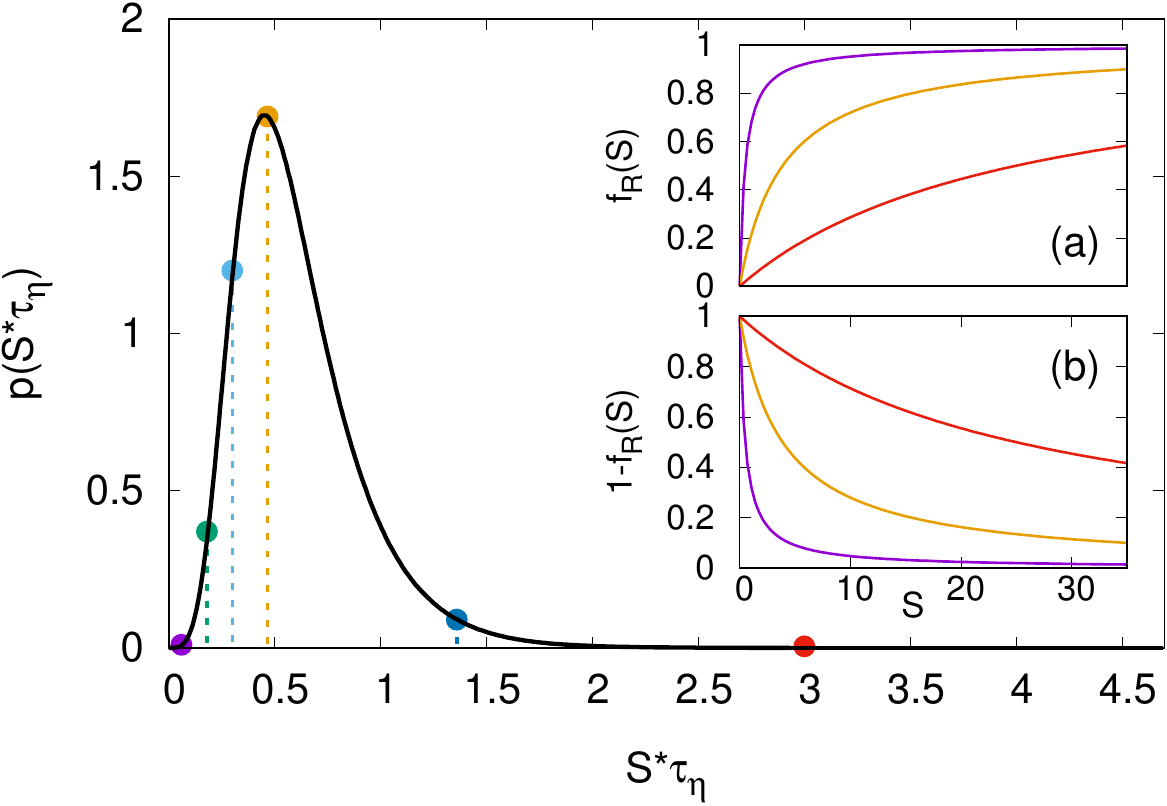}
\caption{Probability density function (PDF) of the Eulerian strain rate. 
Colored dots indicate six different values of $S_{H}$ used in our DNS:
$S_H \tau_{\eta}=0.06, 0.18, 0.3, 0.47, 1.36, 3.0$.
Insets show the response function for shear thickening (a) and for 
shear thinning (b) particles, with different colors corresponding to
the values of $S_H$ indicated in the main panel: $S_{H1} \tau_{\eta}=0.06$ 
(violet), $S_{H4} \tau_{\eta}=0.47$ (orange) and $S_{H6} \tau_{\eta}=3.0$ (red).
Colors online.}
\label{fig:S_eul}
\end{figure}

The numerical populations differ both in the relative sensitivity to the
hydrodynamic cues and in the maximum sedimentation velocity they can reach.
We have chosen $1\leq\Pi\leq30$ (where $\Pi=v_{s,max}/u_\eta$), 
which corresponds to $0.18\leq v_{s,max} \leq 5.4 \, mm/s$ 
when we rescale time and space such that
$\varepsilon = 10^{-9} \, W/kg$, so that we can cover rather well the range of
typical oceanic values of settling speed \cite{Ruiz2004}.
The range of reported values is quite wide, since there is
a significant variation in terms of diatoms species,
methods for analyzing sinking speeds and experimental conditions
(nutrients, light, temperature, flows)
\cite{Gemmell2016,Miklasz2010,Moore1996,Ruiz2004,Villareal1992,Smayda1974,Waite1997}.
In order to choose a
set of values of $S_{H}$ that represent different relevant situations we
computed the Eulerian strain rate of the turbulent flow. 
Numerical results are presented at fixed intermediate turbulence level 
$\varepsilon$, except for sec~\ref{sec:exit_time} where the effects of 
different turbulent intensities on the sedimentation time are discussed.
Figure~\ref{fig:S_eul} shows the probability density function (PDF) of 
$S$; colored dots on the PDF curve indicate the six different values 
of the strain constant $S_H$ used in the simulations of the adaptive 
cells.

\subsection{Clustering and Preferential Sampling} 
\label{sec:clustering}
We first discuss the formation of cell clusters as a result of buoyancy regulation. 
We measure inhomogeneities in the spatial distribution of a population by
its correlation dimension $D_{2}$, defined as the scaling exponent of the probability to find 
two particles at a distance less than 
$r:P(|\bm X_{1} -  \bm X_{2}| < r )\propto r^{D_{2}}$, as $r\rightarrow 0$. $D_2$ 
is directly related to encounter rates between cells, which is a crucial determinant 
for ecological interactions \cite{kiorboe2008}. 
If a given type of interaction happens only once cells are at a certain distance $\bar{r}$, 
the rate at which two cells get close enough for such interaction to occur is proportional to i) the
probability density for the cells to be at exactly that distance $P(r=\bar{r})$, and ii) the typical
relative velocity of cells at that distance \cite{bec2005clustering,falkovich2002acceleration}. 
The former is simply given by the probability density 
for the particles to be on a spherical surface of radius $\bar{r}$, which is $\propto \bar{r}^{D_2-1}$.
In short, if $D_2<3$, the probability of having particles at small distances decays more slowly
as $r\rightarrow 0$ than in a homogeneously distributed population and, as a result, encounter rates increase.

\begin{figure}[htb!]
\centering {\includegraphics[width=\columnwidth]{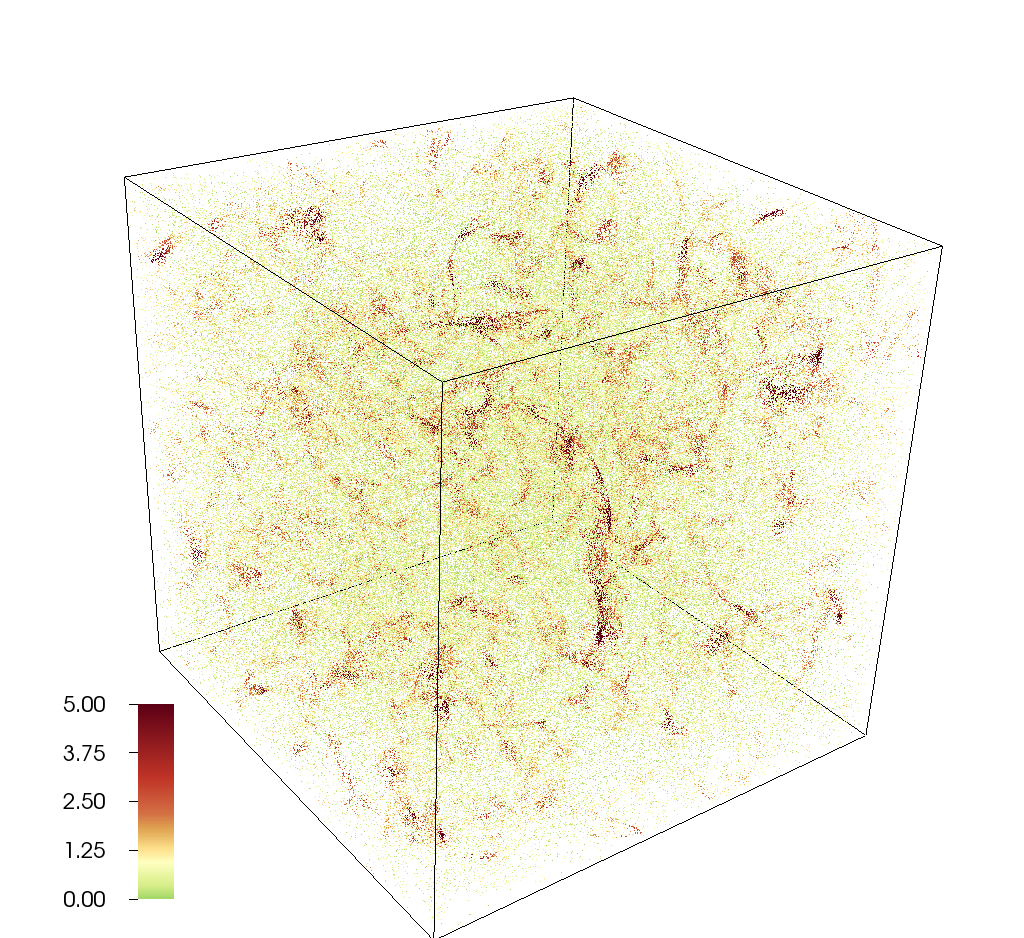}}
\caption{Snapshot of $3 \times 10^5$ thinning particles, with $\Pi=30$, $S_H=S_{H1}$
and $\varepsilon=10^{-9}\, W/kg$.
Each particle is colored according to the ratio of the local number 
density around it to the average density.}
\label{fig:cluster3d}
\end{figure}
	
\begin{figure*}[ht!] 
\includegraphics[width=\textwidth]{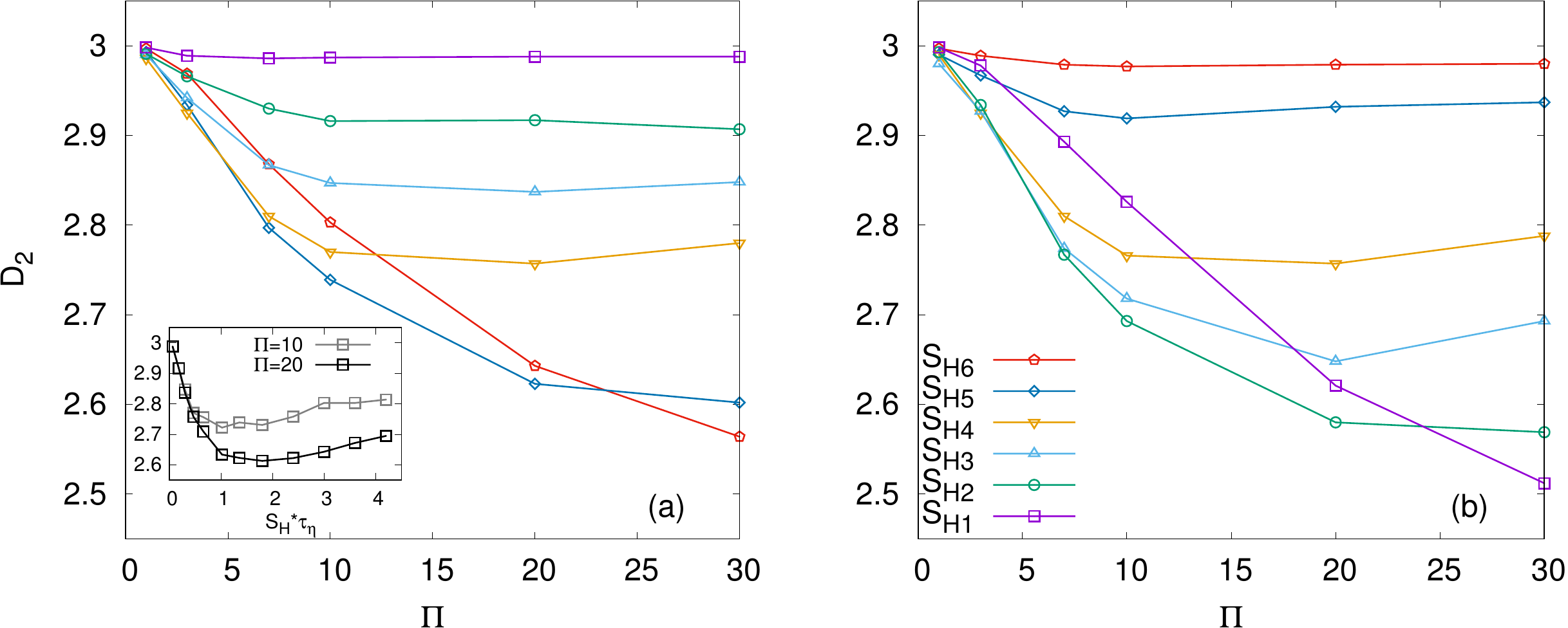}
\caption{Correlation dimension as a function of $\Pi =v_{s,max}/u_{\eta}$ for
shear thickening cells (left) and for shear thinning cells (right).
Inset shows the correlation dimension as a function of the strain rate half
saturation constant in the case of thickening cells with $\Pi=10$ (grey)
and $\Pi=20$ (black).}
\label{fig:d2} 
\end{figure*}

For a homogeneous distribution in a space of dimension $d$ (here $d=3$) 
one has $D_{2} = d$, while $D_{2} < d$
indicates fractal clustering. In the absence of buoyancy regulation, our
simulations show no clustering, as it is expected since in this case particle
velocity is given by a constant downwards term added to an incompressible
velocity field. When regulation is switched on, we observe parameter-dependent
clustering for both shear-thinning and shear-thickening cells 
(see the example in Figure~\ref{fig:cluster3d}). 
In Figure~\ref{fig:d2} we plot the correlation dimension computed 
for both modes of regulation  as a function of $\Pi$ and for different 
values of $S_{H}$. In both
cases when the sedimentation velocity is small we find $D_{2}\simeq 3$,
signaling that distributions remain homogeneous regardless of $S_{H}$. On the
other hand, for large values of $\Pi$ clustering strongly depends on the
response to hydrodynamic stresses.

In the shear thickening case clustering is maximum (i.e., $D_{2}$ is minimum)
for large values of $\Pi$ and $S_{H}$ while $D_{2}\simeq3$ when $S_{H}$ is very
low. In this limit cells tend to become very heavy ($v_{s}\sim v_{s,max}$)
and fast saturation means that correlation with the flow
rapidly become very weak. As a consequence, cells behave very similarly to the
unregulated case and mostly sink with a constant speed. In the opposite limit (large
$S_H$) we have that $v_{s}\sim S \frac{v_{s,max}}{S_{H}}$ and observe
clustering. In this case, regulation is very sensitive and variations in
sedimentation speed are strongly correlated with the flow. 
However, in order to have a relevant degree of clustering, $v_{s,max}$
has to be large to balance the large values of $S_{H}$,
otherwise particles behave as slow sinkers with a consequent
reduction in fractal clustering compared to the other $S_H$ curves
at same value of $\Pi$. The non-monotonic behaviour of $D_2$ with
$S_H$ is shown in the inset of Figure~\ref{fig:d2}
for thickening cells at $\Pi=10$ and $\Pi=20$. 

In similar way shear thinning cells display stronger clustering by increasing
the sedimentation velocity, while the dependency from the strain constant is
opposite compared to the first case. Indeed in the limit of large $S_{H}$ cells
tend to sink since $v_{s}\sim v_{s,max}$ and we find $D_{2}\simeq3$, while for
small $S_{H}$ fractal clustering can take place. The latter limit is not obvious
because it would seem that particles behave like passive tracers ($v_{s}\sim
0$), but from the shear thinning density law we obtain that $v_{s}\sim v_{s,max}
\frac{S_{H}}{S}$, so regulation is a first order effect, although $v_{s,max}$
has to be large to compensate for the small value of strain rate constant
similarly to what discussed about shear thickening cells
for the curve at $S_{H6}$.

\begin{figure}[h!] \centering
{\includegraphics[scale=0.7]{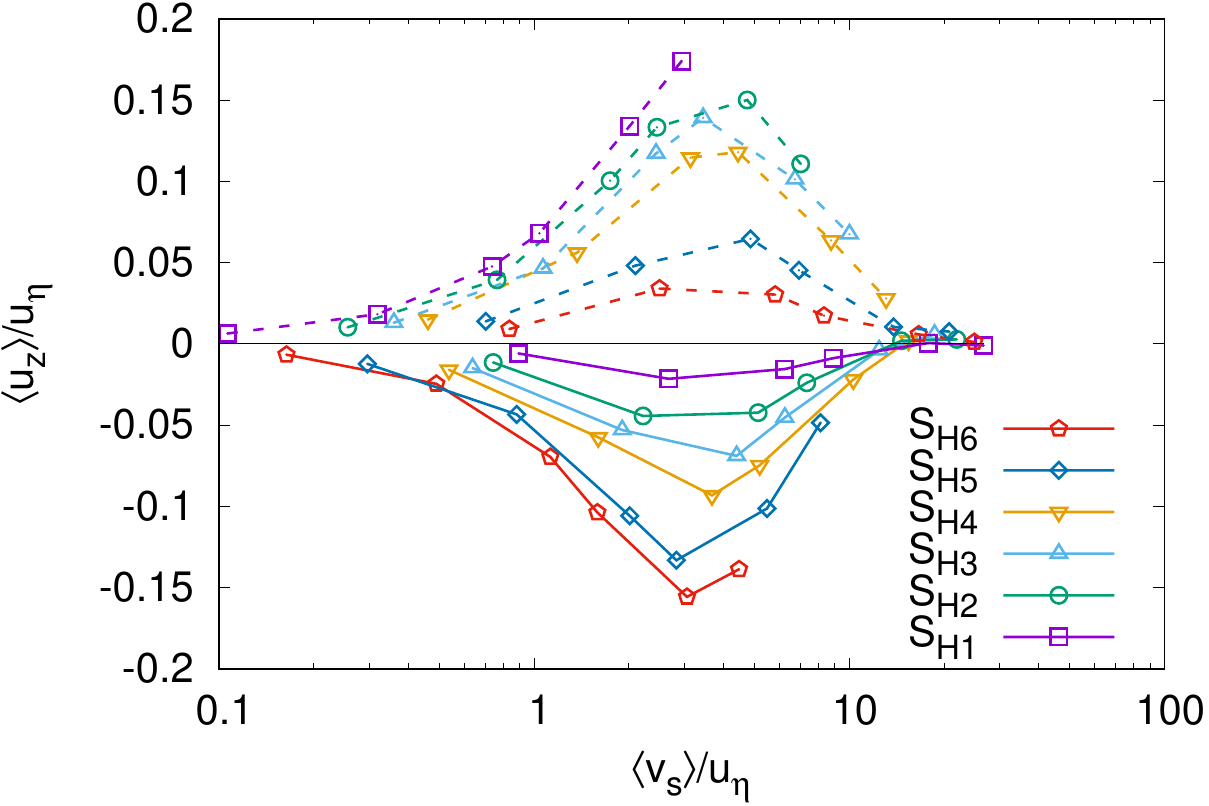}} 
\caption{Average
vertical fluid velocity as a function of mean sedimentation velocity (both
divided by Kolmogorov velocity scale). Different colors represent different
$S_{H}$ while solid (dashed) lines correspond to shear thickening (thinning)
cells.} \label{fig:sampling} \end{figure}

Small scale clustering is often accompanied by a preferential sampling of
regions characterized by certain properties of the flow. This behavior has been
observed both in inertial particles
\cite{Squires1991preferential,Bec2005multifractal,Cencini2006dynamics} and
swimming phytoplankton
\cite{durham2013turbu,Gustavsson2016preferential,Borgnino2018gyrotactic},
and it is also present in this case. 
Figure~\ref{fig:sampling} depicts the average vertical velocity of the fluid,
$\langle u_z \rangle$, calculated on the particle positions as a function of 
the mean sedimentation speed $\langle v_s \rangle$.
While shear-thinning cells appear to spend more
time in regions of upwards flow velocity, shear-thickening particles
preferentially sample downwards velocities. This effect does not affect deeply
the cells' dynamics, since the contribution of the average vertical fluid velocity is small 
compared to the average sedimentation speed,
and it is related to the different buoyancy response.
Indeed in both cases the preferential sampling is larger for the same values of
$S_{H}$ for which the strongest clustering occurs; that means largest shear half
saturation constant for shear thickening particles and lowest $S_{H}$ for
thinning cells.

\subsection{Sedimentation time} 
\label{sec:exit_time}
We explore the effects of buoyancy regulation
on sinking particles by looking at the distribution of sedimentation times,
defined as the time $T_s$ needed to cover a certain vertical distance $L$
which we will take as a multiple of $L_B$.
We compare each population of thinning/thickening cells with passive particles 
sinking at the respective maximum speed $v_{s,max}$ without regulation.

\begin{figure*}[htb!] 
\includegraphics[width=\textwidth]{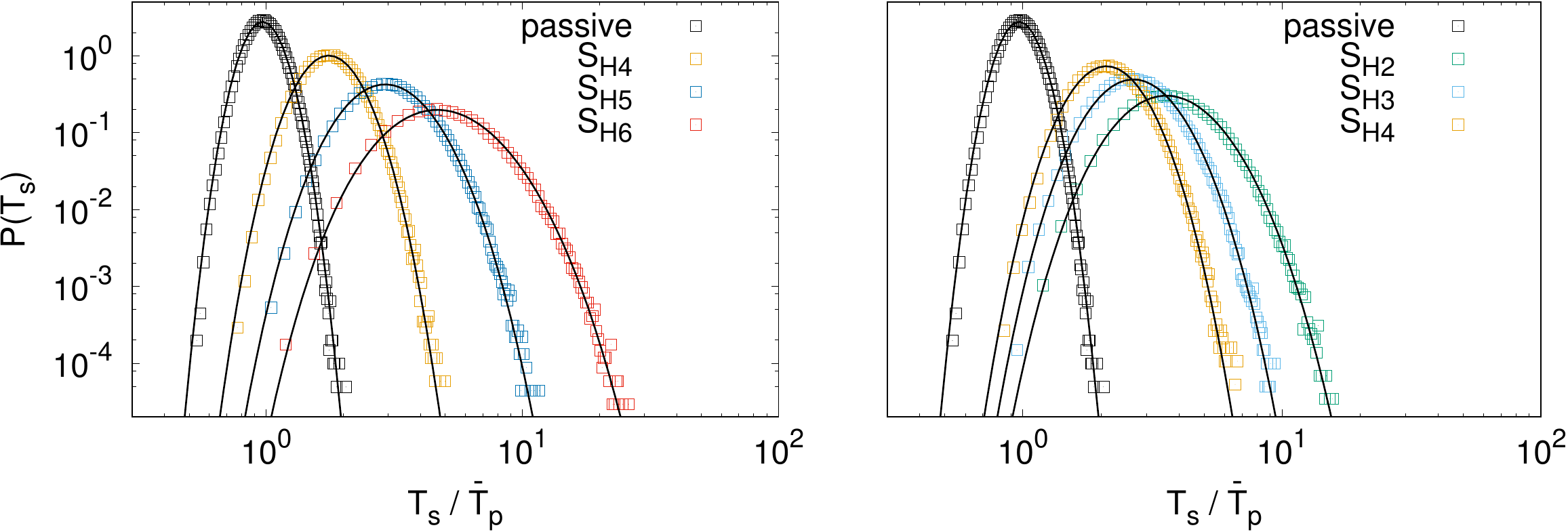}
\caption{Probability distribution of the sedimentation time $T_s$ (divided by
$\overline{T}_p$, the mean sedimentation time of passive cells) for shear thickening cells
(left) and for shear thinning cells (right) log-log scale. 
$\Pi=10$ and $L=4 L_B$ in both cases. 
Black solid lines represent the inverse Gaussian prediction while squares are numerical data. 
Black ones refer to maximum density passive cells, orange to $S_{H4}$, dark blue to $S_{H5}$,
red to $S_{H6}$; while on the right green is $S_{H2}$ and light blue is $S_{H3}$.} 
\label{fig:tau_th}
\end{figure*}

The average time needed by the passive particles to cover the distance
$L$, is known a priori and it is smaller than for active cells.
Figure~\ref{fig:tau_th} shows the comparison between the PDFs of $T_s$ for the
active populations, normalized with the average sedimentation time for 
passive cells $\overline{T}_{p}=L/v_{s,\rm max}$, for both shear thickening 
and thinning cells for the case $\Pi=10$ and for $L=4L_B$. 
It is evident the remarkable difference between the PDFs' maxima: as 
expected, buoyancy control allows cells to sink more slowly 
compared to maximum density passive particles. 
The mean sedimentation time of shear-thickening
particles, increases for larger values of $S_{H}$, while the opposite is
true for shear thinning cells. 
The major effect of buoyancy regulation in this respect is to lower 
the time averaged "instantaneous cell density" (and, hence, the instantaneous sinking speed)
well below the maximum passive value, effectively keeping cells suspended
for much longer times before ultimately sinking to the deep ocean.
Moreover, it is evident that not only the mean sedimentation time increases
(since the comparison is made with unregulated cells with $v_s=v_{s,\rm max}$)
but also the distribution becomes broader, with wider tails in the PDFs. 
In other words, buoyancy regulation does not simply
shift rigidly the PDF of the sedimentation time but also modifies its shape:
this implies that in the shear thickening (thinning) case, for large (small)
$S_H$ a number of cells, contributing to the right tail of the PDF, will remain
suspended for a time significantly larger than the average population.

This behavior can be rationalized by considering the vertical motion 
of sinking particles as a stochastic process with drift. 
Indeed we can replace the deterministic vertical motion described by 
(\ref{eq:2.2}) with a stochastic version with a drift given by the 
average of the vertical component of particle's velocity
$V_d=\langle u_{p,z} \rangle$
and a diffusion coefficient $D_z$ which takes into account 
the turbulent fluctuations (assumed Gaussian), 
including the fluctuations of $S$ which induce the modulations of $v_s$.
In this way, we can recast the problem as a standard first-passage problem 
\cite{Redner2001} where the PDF of the sedimentation times takes 
the form of an inverse Gaussian function \cite{santamaria2014gyrotactic}
(detailed derivation is given in the Supplementary Materials)
\begin{equation}
P(T_s) = \dfrac{L}{(4\pi D_z T_s^3)^{1/2}} e^{-\dfrac{(V_d T_s - L)^2}{4 D_z T_s}}
\label{eq:inv_gaussian}
\end{equation}
In Figure~\ref{fig:tau_th} we plot this analytical predictions with 
the only free parameter, $D_z$, fixed by the value of the variance 
of sedimentation times which, according to (\ref{eq:inv_gaussian}),
is given by $\langle T_s^2 \rangle - \langle T_s \rangle ^2 = 2 D_z L / V_d^3$.
As mentioned above, buoyancy regulation affects the shape
of the PDFs resulting in an increase of $D_z$ (i.e. wider tails of PDFs) 
as a consequence of the turbulent fluctuations that, through $S$ variations,
control $v_s$.
This provides an excellent agreement with the numerical data
which confirms the validity of our approach.

\begin{figure*}[ht!] 
\includegraphics[width=0.9\textwidth]{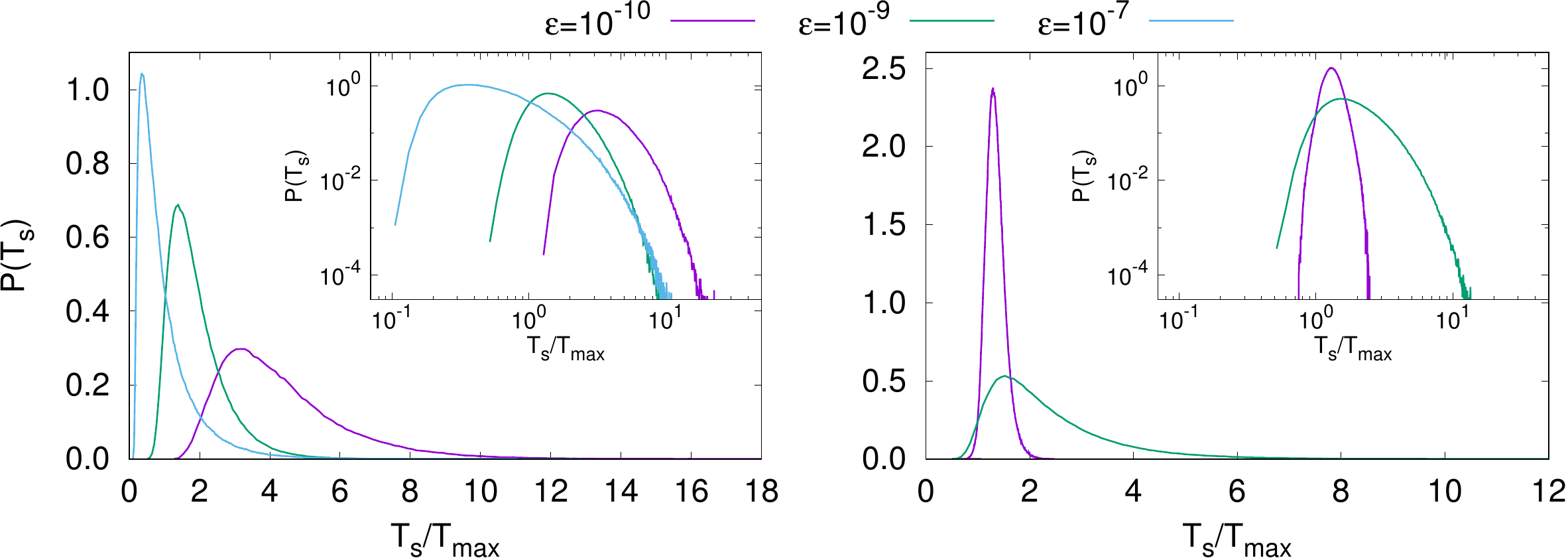}
\caption{Probability distribution of the sedimentation time $T_s$ (divided by
$T_{max}$, the sedimentation time of a particles sinking with $\Pi=10$) for shear thickening cells
(left) and for shear thinning cells (right). Different colors represent different values of 
$\varepsilon$. The insets show the same figures but in log-log scale.} 
\label{fig:re}
\end{figure*}

In a realistic physical situation, phytoplankton cells will face different flow
environments, with possibly vastly different turbulent intensities on seasonal
or even daily basis. It is therefore interesting to study how 
the sedimentation time of a cell with a fixed set of parameters 
depends on the strength of the turbulent flow.
To this aim we performed simulations at three values of energy dissipation rate
$\varepsilon \simeq 10^{-10}$, $10^{-9}$ and $10^{-7}$~W/kg, numerically
obtained by increasing the intensity of the forcing at constant viscosity.
We remark that this scenario is not equivalent to simply taking in consideration
different values of $S_H$, as done in Figures~\ref{fig:S_eul}-\ref{fig:tau_th},
since the statistics of the strain changes with $Re$.
Figure~\ref{fig:re} shows the sedimentation time statistics for cells
with $\Pi=10$ and $S_H=S_{H4}$ in the flows of different intensities.
When turbulence is more intense, particles experience, in general, 
larger values of the local shear. 
One can parameterize such effect, for example, by non-dimensionalizing 
$S_H$ with the the value $S_{peak}$ corresponding to the maximum in the PDF of
strain in each run. The three cases considered have 
$S_H/S_{peak}=3.76$, $1.02$ and $0.07$ respectively.
As a consequence, more intense turbulence
produces faster sedimentation for shear thickening particles.
On the contrary, when shear thinning particles experience intense turbulence,
they become extremely light, almost neutral, leading to large sedimentation
times. 
Since the longest sedimentation times are observed at the largest $Re$,
which is in turn more computationally expensive, we present here only the
statistics for the two lower values of $\varepsilon$ in the shear-thinning
case. For the same reason, sedimentation times are computed for $L=L_B$.
It can be appreciated, from the left panel of Figure~\ref{fig:re}, that
shear-thickening cells develop wider tails as the flow Re increases.
For the case of shear-thinning cells, increased shear slows down the sedimentation. 
However, increased turbulence also widens the distribution of sedimentation times, 
so that many cells sediment faster or slower than the average.
This is probably the most remarkable consequence of this kind of buoyancy regulation 
for diatoms living in a changing turbulent environment: different levels of shear, indeed, 
would not only change the average sedimentation speed, 
but affect the shape of the distribution of sedimentation times. 

\subsection{Segregation} 
Having determined that both shear-thickening and
shear-thinning cells show small scale clustering, we then focus on the relative
spacial distribution of different populations. Community segregation (or the
degree of spatial overlap) is a hallmark of biological diversity as it
facilitates the competition for resources by allowing distinct populations to
explore different ecological niches. Here we look at how community segregation
depends on the clustering properties of the individual populations. We quantify
segregation by the metric defined in \cite{Calzavarini2008}: 
\begin{equation}
S_{1,2}(r)=\dfrac{1}{N_1+N_2} \sum\limits_{i=1}^{M(r)} 
\vert n_{i}^1 - n_{i}^2 \vert 
\label{eq:segreg} 
\end{equation} 
where $M(r)$ is the number of cubes in
which the volume $L^3$ is partitioned (since this approach is based on a coarse
graining over a scale $r$). We computed the segregation between one
shear-thickening and one shear-thinning population characterized by different
parameters $(S_H,\Pi)$. We indicate by $N_1$ and $N_2$ the total number of
particles of each type while $n_i^1$ and $n_i^2$ are the number of particles of
either population contained in each cube $i$. The observable in
eq.~\ref{eq:segreg} varies in the range $\left[0,1\right]$; $S_{1,2}(r)=0$
implies that the total number of the two types is the same at scale $r$ while
the limit $S_{1,2}(r)=1$ means that there is no overlapping between the two
distributions at the considered scale. Complete separation is expected at very
small scale, giving that $\lim_{r \to 0} S_{1,2}(r)=1$, while no structure can
be observed on the scale of the numerical box, with $\lim_{r \to L}
S_{1,2}(r)=0$.
Finally, we define the segregation length scale - the scale up to which the two
distributions do not overlap  - as $R^*=r(S_{1,2}=1/2)$.
At scales below $R^*$, one population is sensibly more abundant than the other 
and, hence, cross-population encounters will be depleted except 
on the boundaries of such areas.

\begin{figure}[h] 
\centering
{\includegraphics[width=0.9\columnwidth]{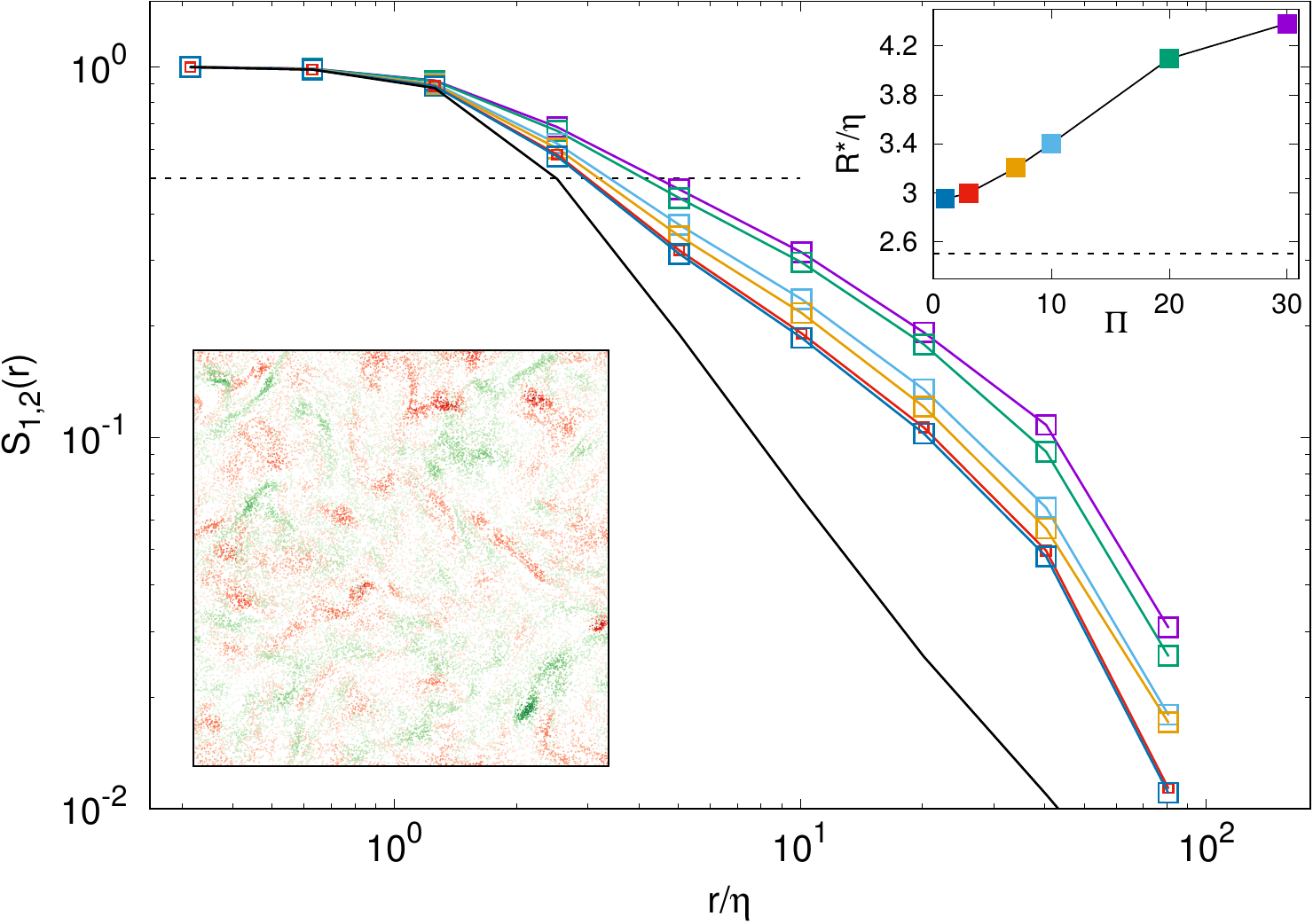}} 
\caption{$S_{1,2}(r)$ curves calculated between a population of shear thinning cells with  
$S_H=S_{H1}$ and $\Pi=30$ and shear thickening cells with constant $S_H=S_{H6}$
and $\Pi=30$ (dark violet), $\Pi=20$ (green), $\Pi=10$ (light blue),
$\Pi=7$ (orange), $\Pi=3$ (red) and $\Pi=1$ (dark blue).
Solid black line corresponds to homogeneously distributed
particles and dashed line to $S_{1,2}(r)=1/2$. 
Bottom Inset: Horizontal section of a typical particle distribution for the clustered case
(dark violet solid line in the main plot); red and green refers respectively to thinning and
thickening particles; in both cases dark/light red (green) indicates high/low cells concentration.
Top inset: Segregation length $R^*$ defined by $S_{1,2}(R^*)=1/2$, as a function $\Pi$, same colors of main plot.
Dashed line represents the value of $R^*$ associated to the case of homogeneously distributed
particles. 
\label{fig:segreg} }
\end{figure}

We consider here both populations with maximum clustering and
having quasi homogeneous distributions. 
For the case of homogeneously distributed particles we take those having
correlation dimension closer to $3$ and sedimentation velocity close to zero, 
representing Poissonian samples.
As examples of inhomogeneous distributions we have considered
the case of the strongest clustering: shear thickening cells characterized by the
largest $S_{H}$ and thinning particles with the smallest strain constant, in
both cases setting $\Pi = 30$. 
Figure~\ref{fig:segreg} shows the comparisons just described.
Furthermore we have compared the values $S_H=S_{H1}$ and $\Pi=30$
for thinning cells with different $\Pi$ for thickening particles
(while the condition $S_H=S_{H6}$ has not changed), in order to
study how the segregation length scale varies with
the parameters.
The top inset of Figure~\ref{fig:segreg} shows the segregation length, 
rescaled by the Kolmogorov length scale, for the different 
parameters' combinations as a function 
of the non-dimensional maximum sedimentation velocity.
The bottom inset of Figure~\ref{fig:segreg} shows the cells 
in a thin slab taken from the numerical box, for the case of maximum clustering
and consequently maximum segregation.  

Although the correlation dimension of the attractors for the two individual 
populations characterized by the maximum clustering 
is, under the chosen set of parameters, almost identical,
their dynamics follow very different rules. As a consequence,
the two attractors are not necessarily overlapping,
leading to well segregated populations as shown in Figure~\ref{fig:segreg}.
It is also possible to appreciate how the correlation length $R^*$ 
reduces accordingly to the decrease of clustering.

\section{Conclusions} \label{sec:conclusions} In this paper we present the first analysis
of the impact of turbulence on the spatial inhomogeneous distribution of
non-motile, but actively responsive, phytoplankton in the ocean. By means of
thorough numerical simulations of a minimal model of active buoyancy regulation
of cells embedded in three-dimensional isotropic turbulence we show that the
non-linear interplay between advection by turbulent flows and cellular activity
leads to cell clustering in low dimensional patches (fractal manifolds), it
affects average sinking rates and it promotes the segregation of distinct
populations. Clustering prompts encounter rates which are key to ecological
processes fundamental for population survival, such as sexual reproduction,
grazing avoidance, or chemical signalling. At the same time, clustering
accelerates physical coagulation mechanisms and, ultimately, the formation of
marine snow, coupling buoyancy control with global biogeochemical cycles
through the regulation of vertical fluxes of organic carbon in the ocean (i.e.
the biological carbon pump). Community segregation, on the contrary,
facilitates the competition for resources by allowing distinct populations to
explore different ecological niches.We also observed a preferential sampling
of certain regions of the flow based on the sign of the vertical component of
fluid velocity. This effect could in principle be relevant in the case of
intense, coherent structures. One such example is Langmuir circulation \cite{Langmuir1938,Thorpe2004}.
In the latter case, the proposed mechanism could lead to accumulation along the
upwelling and downwelling regions between the circulation rolls. The ecological
relevance of inhomogeneous planktonic distributions (which can be produced by many different
dynamics \cite{Barstow1983,Lindemann2017}) along Langmuir circulation 
has been noted by several authors (see \cite{Barstow1983} for a review).

In order to emphasize the significance of active cell
	mechano-responses in spatial inhomogeneities, we have intentionally left out of
	our minimal description any other biological processes affecting population
	dynamics. However, as soon as the characteristic time of sinking becomes of the
	order of the characteristic time of population growth, the interplay
	between purely biological (growth of the population) and physico-biological
	mechanisms (buoyancy control) becomes relevant. This interplay should be
	addressed in future studies. Furthermore, a detailed experimental
characterization of cell responses to mechanical stresses, beyond qualitative
first accounts~\cite{Falciatore2000}, is still pressing. 
In particular, quantifying physiological responses to
hydrodynamic stresses by directly measuring changes in the sinking rate when
cells are exposed to different flow conditions is paramount.  Finally, we would
like to remark once again that a similar, hypothetical
mechanism of buoyancy regulation would likely be much more complex than the
minimal model considered here, as it would depend on many factors, including
environmental conditions and the physiological state of the cell. If buoyancy
regulation similar to the model proposed is found to be realized in
phytoplankton, our analysis would provide a further example of the role played
by turbulence in shaping oceanic community structure not solely through its
large scale direct effect on phytoplankton dispersal but also through less
explored subtle physical-biological coupling at the single cell level.


\section*{Data accessibility}
Scripts and data used to produce figures can be found at:
\href{https://github.com/mborgnino/data-buoyancy-regulating-phytoplankton.git}{https://github.com/mborgnino/data-buoyancy-regulating-phytoplankton.git} 

\section*{Author's Contributions}
All authors contributed equally.

\section*{Funding}
This article is based upon work from COST Action
MP1305, supported by COST (European Cooperation in Science and Technology).
We acknowledge support by the Departments of Excellence grant (MIUR).
HPC Center CINECA is gratefully acknowledged for computing resources,
within the INFN-Cineca agreement INF18-fldturb and IscrC-DnT project.
We acknowledge support from the Spanish Ministry of Economy 
and Competitiveness (AEI, FEDER EU) Grants No. FIS2016-77692-C2-1-P (IT) 
and CTM-2017-83774-D (JA), and the subprogram Juan de la Cierva No. IJCI-2015-26955 (JA).

\begin{acknowledgments}
We thank M. Cencini and G. Dematteis for useful discussions.
\end{acknowledgments}


\begin{thebibliography}{56}%
\makeatletter
\providecommand \@ifxundefined [1]{%
 \@ifx{#1\undefined}
}%
\providecommand \@ifnum [1]{%
 \ifnum #1\expandafter \@firstoftwo
 \else \expandafter \@secondoftwo
 \fi
}%
\providecommand \@ifx [1]{%
 \ifx #1\expandafter \@firstoftwo
 \else \expandafter \@secondoftwo
 \fi
}%
\providecommand \natexlab [1]{#1}%
\providecommand \enquote  [1]{``#1''}%
\providecommand \bibnamefont  [1]{#1}%
\providecommand \bibfnamefont [1]{#1}%
\providecommand \citenamefont [1]{#1}%
\providecommand \href@noop [0]{\@secondoftwo}%
\providecommand \href [0]{\begingroup \@sanitize@url \@href}%
\providecommand \@href[1]{\@@startlink{#1}\@@href}%
\providecommand \@@href[1]{\endgroup#1\@@endlink}%
\providecommand \@sanitize@url [0]{\catcode `\\12\catcode `\$12\catcode
  `\&12\catcode `\#12\catcode `\^12\catcode `\_12\catcode `\%12\relax}%
\providecommand \@@startlink[1]{}%
\providecommand \@@endlink[0]{}%
\providecommand \url  [0]{\begingroup\@sanitize@url \@url }%
\providecommand \@url [1]{\endgroup\@href {#1}{\urlprefix }}%
\providecommand \urlprefix  [0]{URL }%
\providecommand \Eprint [0]{\href }%
\providecommand \doibase [0]{http://dx.doi.org/}%
\providecommand \selectlanguage [0]{\@gobble}%
\providecommand \bibinfo  [0]{\@secondoftwo}%
\providecommand \bibfield  [0]{\@secondoftwo}%
\providecommand \translation [1]{[#1]}%
\providecommand \BibitemOpen [0]{}%
\providecommand \bibitemStop [0]{}%
\providecommand \bibitemNoStop [0]{.\EOS\space}%
\providecommand \EOS [0]{\spacefactor3000\relax}%
\providecommand \BibitemShut  [1]{\csname bibitem#1\endcsname}%
\let\auto@bib@innerbib\@empty
\bibitem [{\citenamefont {Williams}\ and\ \citenamefont
  {Follows}(2011)}]{Williams2011}%
  \BibitemOpen
  \bibfield  {author} {\bibinfo {author} {\bibfnamefont {R.~G.}\ \bibnamefont
  {Williams}}\ and\ \bibinfo {author} {\bibfnamefont {M.~J.}\ \bibnamefont
  {Follows}},\ }\href@noop {} {\emph {\bibinfo {title} {Ocean dynamics and the
  carbon cycle: Principles and mechanisms}}}\ (\bibinfo  {publisher} {Cambridge
  University Press},\ \bibinfo {year} {2011})\BibitemShut {NoStop}%
\bibitem [{\citenamefont {Mitchell}\ \emph {et~al.}(2008)\citenamefont
  {Mitchell}, \citenamefont {Yamazaki}, \citenamefont {Seuront}, \citenamefont
  {Wolk},\ and\ \citenamefont {Li}}]{Mitchell2008}%
  \BibitemOpen
  \bibfield  {author} {\bibinfo {author} {\bibfnamefont {J.}~\bibnamefont
  {Mitchell}}, \bibinfo {author} {\bibfnamefont {H.}~\bibnamefont {Yamazaki}},
  \bibinfo {author} {\bibfnamefont {L.}~\bibnamefont {Seuront}}, \bibinfo
  {author} {\bibfnamefont {F.}~\bibnamefont {Wolk}}, \ and\ \bibinfo {author}
  {\bibfnamefont {H.}~\bibnamefont {Li}},\ }\href@noop {} {\bibfield  {journal}
  {\bibinfo  {journal} {J. Mar. Syst.}\ }\textbf {\bibinfo {volume} {69}},\
  \bibinfo {pages} {247} (\bibinfo {year} {2008})}\BibitemShut {NoStop}%
\bibitem [{\citenamefont {Legendre}\ and\ \citenamefont
  {Fortin}(1989)}]{Legendre1989}%
  \BibitemOpen
  \bibfield  {author} {\bibinfo {author} {\bibfnamefont {P.}~\bibnamefont
  {Legendre}}\ and\ \bibinfo {author} {\bibfnamefont {M.~J.}\ \bibnamefont
  {Fortin}},\ }\href {\doibase 10.1007/BF00048036} {\bibfield  {journal}
  {\bibinfo  {journal} {Vegetatio}\ }\textbf {\bibinfo {volume} {80}},\
  \bibinfo {pages} {107} (\bibinfo {year} {1989})}\BibitemShut {NoStop}%
\bibitem [{\citenamefont {Polis}\ \emph {et~al.}(1997)\citenamefont {Polis},
  \citenamefont {Anderson},\ and\ \citenamefont {Holt}}]{Polis1997}%
  \BibitemOpen
  \bibfield  {author} {\bibinfo {author} {\bibfnamefont {G.~A.}\ \bibnamefont
  {Polis}}, \bibinfo {author} {\bibfnamefont {W.~B.}\ \bibnamefont {Anderson}},
  \ and\ \bibinfo {author} {\bibfnamefont {R.~D.}\ \bibnamefont {Holt}},\
  }\href {\doibase 10.1146/annurev.ecolsys.28.1.289} {\bibfield  {journal}
  {\bibinfo  {journal} {Annual Review of Ecology and Systematics}\ }\textbf
  {\bibinfo {volume} {28}},\ \bibinfo {pages} {289} (\bibinfo {year}
  {1997})}\BibitemShut {NoStop}%
\bibitem [{\citenamefont {Mackas}\ \emph {et~al.}(1985)\citenamefont {Mackas},
  \citenamefont {Denman},\ and\ \citenamefont {Abbott}}]{Mackas1985}%
  \BibitemOpen
  \bibfield  {author} {\bibinfo {author} {\bibfnamefont {D.~L.}\ \bibnamefont
  {Mackas}}, \bibinfo {author} {\bibfnamefont {K.~L.}\ \bibnamefont {Denman}},
  \ and\ \bibinfo {author} {\bibfnamefont {M.~R.}\ \bibnamefont {Abbott}},\
  }\href@noop {} {\bibfield  {journal} {\bibinfo  {journal} {Bull. Mar.
  Science}\ }\textbf {\bibinfo {volume} {37}},\ \bibinfo {pages} {652}
  (\bibinfo {year} {1985})}\BibitemShut {NoStop}%
\bibitem [{\citenamefont {Martin}(2003)}]{Martin2003}%
  \BibitemOpen
  \bibfield  {author} {\bibinfo {author} {\bibfnamefont {A.}~\bibnamefont
  {Martin}},\ }\href@noop {} {\bibfield  {journal} {\bibinfo  {journal} {Progr.
  Ocean.}\ }\textbf {\bibinfo {volume} {57}},\ \bibinfo {pages} {125} (\bibinfo
  {year} {2003})}\BibitemShut {NoStop}%
\bibitem [{\citenamefont {Ki{\o}rboe}(2008)}]{kiorboe2008}%
  \BibitemOpen
  \bibfield  {author} {\bibinfo {author} {\bibfnamefont {T.}~\bibnamefont
  {Ki{\o}rboe}},\ }\href@noop {} {\emph {\bibinfo {title} {A mechanistic
  approach to plankton ecology}}}\ (\bibinfo  {publisher} {Princeton University
  Press},\ \bibinfo {year} {2008})\BibitemShut {NoStop}%
\bibitem [{\citenamefont {Visser}\ and\ \citenamefont
  {Ki{\o}rboe}(2006)}]{Visser2006}%
  \BibitemOpen
  \bibfield  {author} {\bibinfo {author} {\bibfnamefont {A.~W.}\ \bibnamefont
  {Visser}}\ and\ \bibinfo {author} {\bibfnamefont {T.}~\bibnamefont
  {Ki{\o}rboe}},\ }\href@noop {} {\bibfield  {journal} {\bibinfo  {journal}
  {Oecologia}\ }\textbf {\bibinfo {volume} {148}},\ \bibinfo {pages} {538}
  (\bibinfo {year} {2006})}\BibitemShut {NoStop}%
\bibitem [{\citenamefont {Durham}\ and\ \citenamefont
  {Stocker}(2012)}]{Durham2012}%
  \BibitemOpen
  \bibfield  {author} {\bibinfo {author} {\bibfnamefont {W.~M.}\ \bibnamefont
  {Durham}}\ and\ \bibinfo {author} {\bibfnamefont {R.}~\bibnamefont
  {Stocker}},\ }\href@noop {} {\bibfield  {journal} {\bibinfo  {journal} {Annu.
  Rev. Mar. Sci.}\ }\textbf {\bibinfo {volume} {4}},\ \bibinfo {pages} {177}
  (\bibinfo {year} {2012})}\BibitemShut {NoStop}%
\bibitem [{\citenamefont {Durham}\ \emph
  {et~al.}(2013{\natexlab{a}})\citenamefont {Durham}, \citenamefont {Climent},
  \citenamefont {Barry}, \citenamefont {{De Lillo}}, \citenamefont {Boffetta},
  \citenamefont {Cencini},\ and\ \citenamefont {Stocker}}]{Durham2013}%
  \BibitemOpen
  \bibfield  {author} {\bibinfo {author} {\bibfnamefont {W.~M.}\ \bibnamefont
  {Durham}}, \bibinfo {author} {\bibfnamefont {E.}~\bibnamefont {Climent}},
  \bibinfo {author} {\bibfnamefont {M.}~\bibnamefont {Barry}}, \bibinfo
  {author} {\bibfnamefont {F.}~\bibnamefont {{De Lillo}}}, \bibinfo {author}
  {\bibfnamefont {G.}~\bibnamefont {Boffetta}}, \bibinfo {author}
  {\bibfnamefont {M.}~\bibnamefont {Cencini}}, \ and\ \bibinfo {author}
  {\bibfnamefont {R.}~\bibnamefont {Stocker}},\ }\href {\doibase
  10.1038/ncomms3148} {\bibfield  {journal} {\bibinfo  {journal} {Nat. Comm.}\
  }\textbf {\bibinfo {volume} {4}},\ \bibinfo {pages} {2148} (\bibinfo {year}
  {2013}{\natexlab{a}})}\BibitemShut {NoStop}%
\bibitem [{\citenamefont {Margalef}(1978)}]{Margalef1978}%
  \BibitemOpen
  \bibfield  {author} {\bibinfo {author} {\bibfnamefont {R.}~\bibnamefont
  {Margalef}},\ }\href@noop {} {\bibfield  {journal} {\bibinfo  {journal}
  {Oceanologica Acta}\ }\textbf {\bibinfo {volume} {1}},\ \bibinfo {pages}
  {493} (\bibinfo {year} {1978})}\BibitemShut {NoStop}%
\bibitem [{\citenamefont {Azam}\ and\ \citenamefont
  {Malfatti}(2007)}]{Azam2007}%
  \BibitemOpen
  \bibfield  {author} {\bibinfo {author} {\bibfnamefont {F.}~\bibnamefont
  {Azam}}\ and\ \bibinfo {author} {\bibfnamefont {F.}~\bibnamefont
  {Malfatti}},\ }\href {https://doi.org/10.1038/nrmicro1747} {\bibfield
  {journal} {\bibinfo  {journal} {Nature Reviews Microbiology}\ }\textbf
  {\bibinfo {volume} {5}},\ \bibinfo {pages} {782 EP } (\bibinfo {year}
  {2007})},\ \bibinfo {note} {review Article}\BibitemShut {NoStop}%
\bibitem [{\citenamefont {Seymour}\ \emph {et~al.}(2017)\citenamefont
  {Seymour}, \citenamefont {Amin}, \citenamefont {Raina},\ and\ \citenamefont
  {Stocker}}]{Seymour2017}%
  \BibitemOpen
  \bibfield  {author} {\bibinfo {author} {\bibfnamefont {J.~R.}\ \bibnamefont
  {Seymour}}, \bibinfo {author} {\bibfnamefont {S.~A.}\ \bibnamefont {Amin}},
  \bibinfo {author} {\bibfnamefont {J.-B.}\ \bibnamefont {Raina}}, \ and\
  \bibinfo {author} {\bibfnamefont {R.}~\bibnamefont {Stocker}},\ }\href
  {https://doi.org/10.1038/nmicrobiol.2017.65} {\bibfield  {journal} {\bibinfo
  {journal} {Nature Microbiology}\ }\textbf {\bibinfo {volume} {2}},\ \bibinfo
  {pages} {17065 EP } (\bibinfo {year} {2017})},\ \bibinfo {note} {review
  Article}\BibitemShut {NoStop}%
\bibitem [{\citenamefont {Durham}\ \emph
  {et~al.}(2013{\natexlab{b}})\citenamefont {Durham}, \citenamefont {Climent},
  \citenamefont {Barry}, \citenamefont {De~Lillo}, \citenamefont {Boffetta},
  \citenamefont {Cencini},\ and\ \citenamefont {Stocker}}]{durham2013turbu}%
  \BibitemOpen
  \bibfield  {author} {\bibinfo {author} {\bibfnamefont {W.~M.}\ \bibnamefont
  {Durham}}, \bibinfo {author} {\bibfnamefont {E.}~\bibnamefont {Climent}},
  \bibinfo {author} {\bibfnamefont {M.}~\bibnamefont {Barry}}, \bibinfo
  {author} {\bibfnamefont {F.}~\bibnamefont {De~Lillo}}, \bibinfo {author}
  {\bibfnamefont {G.}~\bibnamefont {Boffetta}}, \bibinfo {author}
  {\bibfnamefont {M.}~\bibnamefont {Cencini}}, \ and\ \bibinfo {author}
  {\bibfnamefont {R.}~\bibnamefont {Stocker}},\ }\href@noop {} {\bibfield
  {journal} {\bibinfo  {journal} {Nature Comm.}\ }\textbf {\bibinfo {volume}
  {4}},\ \bibinfo {pages} {2148} (\bibinfo {year}
  {2013}{\natexlab{b}})}\BibitemShut {NoStop}%
\bibitem [{\citenamefont {Berman-Frank}\ \emph {et~al.}(2003)\citenamefont
  {Berman-Frank}, \citenamefont {Lundgren},\ and\ \citenamefont
  {Falkowski}}]{Berman-Frank2003}%
  \BibitemOpen
  \bibfield  {author} {\bibinfo {author} {\bibfnamefont {I.}~\bibnamefont
  {Berman-Frank}}, \bibinfo {author} {\bibfnamefont {P.}~\bibnamefont
  {Lundgren}}, \ and\ \bibinfo {author} {\bibfnamefont {P.}~\bibnamefont
  {Falkowski}},\ }\href {\doibase 10.1016/S0923-2508(03)00029-9} {\bibfield
  {journal} {\bibinfo  {journal} {Research in Microbiology}\ }\textbf {\bibinfo
  {volume} {154}},\ \bibinfo {pages} {157} (\bibinfo {year}
  {2003})}\BibitemShut {NoStop}%
\bibitem [{\citenamefont {Zehr}(2010)}]{Zehr2010}%
  \BibitemOpen
  \bibfield  {author} {\bibinfo {author} {\bibfnamefont {J.~P.}\ \bibnamefont
  {Zehr}},\ }\href {\doibase 10.1016/j.tim.2010.12.004} {\bibfield  {journal}
  {\bibinfo  {journal} {Trends in Microbiology}\ }\textbf {\bibinfo {volume}
  {19}},\ \bibinfo {pages} {162} (\bibinfo {year} {2010})}\BibitemShut
  {NoStop}%
\bibitem [{\citenamefont {Malviya}\ \emph {et~al.}(2016)\citenamefont
  {Malviya}, \citenamefont {Scalco}, \citenamefont {Audic}, \citenamefont
  {Vincent}, \citenamefont {Veluchamy}, \citenamefont {Poulain}, \citenamefont
  {Wincker}, \citenamefont {Iudicone}, \citenamefont {de~Vargas}, \citenamefont
  {Bittner}, \citenamefont {Zingone},\ and\ \citenamefont
  {Bowler}}]{Malviya2016}%
  \BibitemOpen
  \bibfield  {author} {\bibinfo {author} {\bibfnamefont {S.}~\bibnamefont
  {Malviya}}, \bibinfo {author} {\bibfnamefont {E.}~\bibnamefont {Scalco}},
  \bibinfo {author} {\bibfnamefont {S.}~\bibnamefont {Audic}}, \bibinfo
  {author} {\bibfnamefont {F.}~\bibnamefont {Vincent}}, \bibinfo {author}
  {\bibfnamefont {A.}~\bibnamefont {Veluchamy}}, \bibinfo {author}
  {\bibfnamefont {J.}~\bibnamefont {Poulain}}, \bibinfo {author} {\bibfnamefont
  {P.}~\bibnamefont {Wincker}}, \bibinfo {author} {\bibfnamefont
  {D.}~\bibnamefont {Iudicone}}, \bibinfo {author} {\bibfnamefont
  {C.}~\bibnamefont {de~Vargas}}, \bibinfo {author} {\bibfnamefont
  {L.}~\bibnamefont {Bittner}}, \bibinfo {author} {\bibfnamefont
  {A.}~\bibnamefont {Zingone}}, \ and\ \bibinfo {author} {\bibfnamefont
  {C.}~\bibnamefont {Bowler}},\ }\href {\doibase 10.1073/pnas.1509523113}
  {\bibfield  {journal} {\bibinfo  {journal} {Proceedings of the National
  Academy of Sciences}\ }\textbf {\bibinfo {volume} {113}},\ \bibinfo {pages}
  {E1516} (\bibinfo {year} {2016})}\BibitemShut {NoStop}%
\bibitem [{\citenamefont {Gross}\ \emph {et~al.}(1948)\citenamefont {Gross},
  \citenamefont {Zeuthen},\ and\ \citenamefont {Yonge}}]{Gross1948}%
  \BibitemOpen
  \bibfield  {author} {\bibinfo {author} {\bibfnamefont {F.}~\bibnamefont
  {Gross}}, \bibinfo {author} {\bibfnamefont {E.}~\bibnamefont {Zeuthen}}, \
  and\ \bibinfo {author} {\bibfnamefont {M.}~\bibnamefont {Yonge}},\ }\href
  {\doibase 10.1098/rspb.1948.0017} {\bibfield  {journal} {\bibinfo  {journal}
  {Proceedings of the Royal Society of London. Series B - Biological Sciences}\
  }\textbf {\bibinfo {volume} {135}},\ \bibinfo {pages} {382} (\bibinfo {year}
  {1948})}\BibitemShut {NoStop}%
\bibitem [{\citenamefont {Waite}(1992)}]{Waite1992}%
  \BibitemOpen
  \bibfield  {author} {\bibinfo {author} {\bibfnamefont {A.~M.}\ \bibnamefont
  {Waite}},\ }\emph {\bibinfo {title} {Physiological control of diatom
  sedimentation}},\ \href@noop {} {Ph.D. thesis},\ \bibinfo  {school}
  {University of Brithis Columbia} (\bibinfo {year} {1992})\BibitemShut
  {NoStop}%
\bibitem [{\citenamefont {Walsby}(1994)}]{Walsby94}%
  \BibitemOpen
  \bibfield  {author} {\bibinfo {author} {\bibfnamefont {A.~E.}\ \bibnamefont
  {Walsby}},\ }\href@noop {} {\bibfield  {journal} {\bibinfo  {journal}
  {Microbiology and Molecular Biology Reviews}\ }\textbf {\bibinfo {volume}
  {58}},\ \bibinfo {pages} {94} (\bibinfo {year} {1994})}\BibitemShut {NoStop}%
\bibitem [{\citenamefont {Arrieta}\ \emph {et~al.}(2015)\citenamefont
  {Arrieta}, \citenamefont {Barreira},\ and\ \citenamefont
  {Tuval}}]{arrieta2015microscale}%
  \BibitemOpen
  \bibfield  {author} {\bibinfo {author} {\bibfnamefont {J.}~\bibnamefont
  {Arrieta}}, \bibinfo {author} {\bibfnamefont {A.}~\bibnamefont {Barreira}}, \
  and\ \bibinfo {author} {\bibfnamefont {I.}~\bibnamefont {Tuval}},\
  }\href@noop {} {\bibfield  {journal} {\bibinfo  {journal} {Phys. Rev. Lett.}\
  }\textbf {\bibinfo {volume} {114}},\ \bibinfo {pages} {128102} (\bibinfo
  {year} {2015})}\BibitemShut {NoStop}%
\bibitem [{\citenamefont {Falciatore}\ \emph {et~al.}(2000)\citenamefont
  {Falciatore}, \citenamefont {d'Alcalà}, \citenamefont {Croot},\ and\
  \citenamefont {Bowler}}]{Falciatore2000}%
  \BibitemOpen
  \bibfield  {author} {\bibinfo {author} {\bibfnamefont {A.}~\bibnamefont
  {Falciatore}}, \bibinfo {author} {\bibfnamefont {M.~R.}\ \bibnamefont
  {d'Alcalà}}, \bibinfo {author} {\bibfnamefont {P.}~\bibnamefont {Croot}}, \
  and\ \bibinfo {author} {\bibfnamefont {C.}~\bibnamefont {Bowler}},\
  }\href@noop {} {\bibfield  {journal} {\bibinfo  {journal} {Science}\ }\textbf
  {\bibinfo {volume} {288}},\ \bibinfo {pages} {2363} (\bibinfo {year}
  {2000})}\BibitemShut {NoStop}%
\bibitem [{\citenamefont {Falciatore}\ and\ \citenamefont
  {Bowler}(2002)}]{Falciatore2002}%
  \BibitemOpen
  \bibfield  {author} {\bibinfo {author} {\bibfnamefont {A.}~\bibnamefont
  {Falciatore}}\ and\ \bibinfo {author} {\bibfnamefont {C.}~\bibnamefont
  {Bowler}},\ }\href@noop {} {\bibfield  {journal} {\bibinfo  {journal} {Annual
  review of plant biology}\ }\textbf {\bibinfo {volume} {53}},\ \bibinfo
  {pages} {109} (\bibinfo {year} {2002})}\BibitemShut {NoStop}%
\bibitem [{\citenamefont {Erga}\ \emph {et~al.}(2010)\citenamefont {Erga},
  \citenamefont {Lie}, \citenamefont {Aarø}, \citenamefont {Aursland},
  \citenamefont {Olseng}, \citenamefont {Øyvind Frette},\ and\ \citenamefont
  {Hamre}}]{Erga2009}%
  \BibitemOpen
  \bibfield  {author} {\bibinfo {author} {\bibfnamefont {S.~R.}\ \bibnamefont
  {Erga}}, \bibinfo {author} {\bibfnamefont {G.~C.}\ \bibnamefont {Lie}},
  \bibinfo {author} {\bibfnamefont {L.~H.}\ \bibnamefont {Aarø}}, \bibinfo
  {author} {\bibfnamefont {K.}~\bibnamefont {Aursland}}, \bibinfo {author}
  {\bibfnamefont {C.~D.}\ \bibnamefont {Olseng}}, \bibinfo {author}
  {\bibnamefont {Øyvind Frette}}, \ and\ \bibinfo {author} {\bibfnamefont
  {B.}~\bibnamefont {Hamre}},\ }\href@noop {} {\bibfield  {journal} {\bibinfo
  {journal} {Journal of Experimental Marine Biology and Ecology}\ }\textbf
  {\bibinfo {volume} {384}},\ \bibinfo {pages} {7 } (\bibinfo {year}
  {2010})}\BibitemShut {NoStop}%
\bibitem [{\citenamefont {Gemmell}\ \emph {et~al.}(2016)\citenamefont
  {Gemmell}, \citenamefont {Oh}, \citenamefont {Buskey},\ and\ \citenamefont
  {Villareal}}]{Gemmell2016}%
  \BibitemOpen
  \bibfield  {author} {\bibinfo {author} {\bibfnamefont {B.~J.}\ \bibnamefont
  {Gemmell}}, \bibinfo {author} {\bibfnamefont {G.}~\bibnamefont {Oh}},
  \bibinfo {author} {\bibfnamefont {E.~J.}\ \bibnamefont {Buskey}}, \ and\
  \bibinfo {author} {\bibfnamefont {T.~A.}\ \bibnamefont {Villareal}},\
  }\href@noop {} {\bibfield  {journal} {\bibinfo  {journal} {Proc. R. Soc. B}\
  }\textbf {\bibinfo {volume} {283}},\ \bibinfo {pages} {20161126} (\bibinfo
  {year} {2016})}\BibitemShut {NoStop}%
\bibitem [{\citenamefont {Amato}\ \emph {et~al.}(2017)\citenamefont {Amato},
  \citenamefont {Dell’Aquila}, \citenamefont {Musacchia}, \citenamefont
  {Annunziata}, \citenamefont {Ugarte}, \citenamefont {Maillet}, \citenamefont
  {Carbone}, \citenamefont {d’Alcalà}, \citenamefont {Sanges},\ and\
  \citenamefont {Iudicone}}]{Amato2017}%
  \BibitemOpen
  \bibfield  {author} {\bibinfo {author} {\bibfnamefont {A.}~\bibnamefont
  {Amato}}, \bibinfo {author} {\bibfnamefont {G.}~\bibnamefont
  {Dell’Aquila}}, \bibinfo {author} {\bibfnamefont {F.}~\bibnamefont
  {Musacchia}}, \bibinfo {author} {\bibfnamefont {R.}~\bibnamefont
  {Annunziata}}, \bibinfo {author} {\bibfnamefont {A.}~\bibnamefont {Ugarte}},
  \bibinfo {author} {\bibfnamefont {N.}~\bibnamefont {Maillet}}, \bibinfo
  {author} {\bibfnamefont {A.}~\bibnamefont {Carbone}}, \bibinfo {author}
  {\bibfnamefont {M.~R.}\ \bibnamefont {d’Alcalà}}, \bibinfo {author}
  {\bibfnamefont {R.}~\bibnamefont {Sanges}}, \ and\ \bibinfo {author}
  {\bibfnamefont {D.}~\bibnamefont {Iudicone}},\ }\href@noop {} {\bibfield
  {journal} {\bibinfo  {journal} {Scientific Reports}\ }\textbf {\bibinfo
  {volume} {7}},\ \bibinfo {pages} {3826} (\bibinfo {year} {2017})}\BibitemShut
  {NoStop}%
\bibitem [{\citenamefont {Frisch}(1995)}]{frisch1995turbulence}%
  \BibitemOpen
  \bibfield  {author} {\bibinfo {author} {\bibfnamefont {U.}~\bibnamefont
  {Frisch}},\ }\href@noop {} {\emph {\bibinfo {title} {Turbulence: the legacy
  of AN Kolmogorov}}}\ (\bibinfo  {publisher} {Cambridge university press},\
  \bibinfo {year} {1995})\BibitemShut {NoStop}%
\bibitem [{\citenamefont {Maxey}\ and\ \citenamefont
  {Riley}(1983)}]{Maxey1983}%
  \BibitemOpen
  \bibfield  {author} {\bibinfo {author} {\bibfnamefont {M.~R.}\ \bibnamefont
  {Maxey}}\ and\ \bibinfo {author} {\bibfnamefont {J.~J.}\ \bibnamefont
  {Riley}},\ }\href@noop {} {\bibfield  {journal} {\bibinfo  {journal} {The
  Physics of Fluids}\ }\textbf {\bibinfo {volume} {26}},\ \bibinfo {pages}
  {883} (\bibinfo {year} {1983})}\BibitemShut {NoStop}%
\bibitem [{\citenamefont {Provenzale}(1999)}]{Provenzale1999}%
  \BibitemOpen
  \bibfield  {author} {\bibinfo {author} {\bibfnamefont {A.}~\bibnamefont
  {Provenzale}},\ }\href {\doibase 10.1146/annurev.fluid.31.1.55} {\bibfield
  {journal} {\bibinfo  {journal} {Annual Review of Fluid Mechanics}\ }\textbf
  {\bibinfo {volume} {31}},\ \bibinfo {pages} {55} (\bibinfo {year}
  {1999})}\BibitemShut {NoStop}%
\bibitem [{\citenamefont {Babiano}\ \emph {et~al.}(2000)\citenamefont
  {Babiano}, \citenamefont {Cartwright}, \citenamefont {Piro},\ and\
  \citenamefont {Provenzale}}]{Babiano2000}%
  \BibitemOpen
  \bibfield  {author} {\bibinfo {author} {\bibfnamefont {A.}~\bibnamefont
  {Babiano}}, \bibinfo {author} {\bibfnamefont {J.~H.~E.}\ \bibnamefont
  {Cartwright}}, \bibinfo {author} {\bibfnamefont {O.}~\bibnamefont {Piro}}, \
  and\ \bibinfo {author} {\bibfnamefont {A.}~\bibnamefont {Provenzale}},\
  }\href {\doibase 10.1103/PhysRevLett.84.5764} {\bibfield  {journal} {\bibinfo
   {journal} {Phys. Rev. Lett.}\ }\textbf {\bibinfo {volume} {84}},\ \bibinfo
  {pages} {5764} (\bibinfo {year} {2000})}\BibitemShut {NoStop}%
\bibitem [{\citenamefont {Daitche}\ and\ \citenamefont
  {T{\'{e}}l}(2014)}]{Daitche_2014}%
  \BibitemOpen
  \bibfield  {author} {\bibinfo {author} {\bibfnamefont {A.}~\bibnamefont
  {Daitche}}\ and\ \bibinfo {author} {\bibfnamefont {T.}~\bibnamefont
  {T{\'{e}}l}},\ }\href {\doibase 10.1088/1367-2630/16/7/073008} {\bibfield
  {journal} {\bibinfo  {journal} {New Journal of Physics}\ }\textbf {\bibinfo
  {volume} {16}},\ \bibinfo {pages} {073008} (\bibinfo {year}
  {2014})}\BibitemShut {NoStop}%
\bibitem [{\citenamefont {Reynolds}(2006)}]{Reynolds2006}%
  \BibitemOpen
  \bibfield  {author} {\bibinfo {author} {\bibfnamefont {C.~S.}\ \bibnamefont
  {Reynolds}},\ }\href@noop {} {\emph {\bibinfo {title} {The ecology of
  phytoplankton}}}\ (\bibinfo  {publisher} {Cambridge University Press},\
  \bibinfo {year} {2006})\BibitemShut {NoStop}%
\bibitem [{\citenamefont {Thorpe}(2007)}]{Thorpe2007}%
  \BibitemOpen
  \bibfield  {author} {\bibinfo {author} {\bibfnamefont {S.~A.}\ \bibnamefont
  {Thorpe}},\ }\href@noop {} {\emph {\bibinfo {title} {An introduction to ocean
  turbulence}}}\ (\bibinfo  {publisher} {Cambridge University Press},\ \bibinfo
  {year} {2007})\BibitemShut {NoStop}%
\bibitem [{\citenamefont {Ott}(2002)}]{Ott2002}%
  \BibitemOpen
  \bibfield  {author} {\bibinfo {author} {\bibfnamefont {E.}~\bibnamefont
  {Ott}},\ }\href@noop {} {\emph {\bibinfo {title} {Chaos in dynamical
  systems}}}\ (\bibinfo  {publisher} {Cambridge university press},\ \bibinfo
  {year} {2002})\BibitemShut {NoStop}%
\bibitem [{\citenamefont {Sengupta}\ \emph {et~al.}(2017)\citenamefont
  {Sengupta}, \citenamefont {Carrara},\ and\ \citenamefont
  {Stocker}}]{Sengupta2017}%
  \BibitemOpen
  \bibfield  {author} {\bibinfo {author} {\bibfnamefont {A.}~\bibnamefont
  {Sengupta}}, \bibinfo {author} {\bibfnamefont {F.}~\bibnamefont {Carrara}}, \
  and\ \bibinfo {author} {\bibfnamefont {R.}~\bibnamefont {Stocker}},\ }\href
  {\doibase 10.1038/nature21415} {\bibfield  {journal} {\bibinfo  {journal}
  {Nature}\ }\textbf {\bibinfo {volume} {543}},\ \bibinfo {pages} {555}
  (\bibinfo {year} {2017})}\BibitemShut {NoStop}%
\bibitem [{\citenamefont {Boyd}(2001)}]{Boyd2001}%
  \BibitemOpen
  \bibfield  {author} {\bibinfo {author} {\bibfnamefont {J.~P.}\ \bibnamefont
  {Boyd}},\ }\href@noop {} {\emph {\bibinfo {title} {Chebyshev and Fourier
  spectral methods}}}\ (\bibinfo  {publisher} {Courier Corporation},\ \bibinfo
  {year} {2001})\BibitemShut {NoStop}%
\bibitem [{\citenamefont {Ruiz}\ \emph {et~al.}(2004)\citenamefont {Ruiz},
  \citenamefont {Mac\'ias},\ and\ \citenamefont {Peters}}]{Ruiz2004}%
  \BibitemOpen
  \bibfield  {author} {\bibinfo {author} {\bibfnamefont {J.}~\bibnamefont
  {Ruiz}}, \bibinfo {author} {\bibfnamefont {D.}~\bibnamefont {Mac\'ias}}, \
  and\ \bibinfo {author} {\bibfnamefont {F.}~\bibnamefont {Peters}},\
  }\href@noop {} {\bibfield  {journal} {\bibinfo  {journal} {Proceedings of the
  National Academy of Sciences}\ }\textbf {\bibinfo {volume} {101}},\ \bibinfo
  {pages} {17720} (\bibinfo {year} {2004})}\BibitemShut {NoStop}%
\bibitem [{\citenamefont {Miklasz}\ and\ \citenamefont
  {Denny}(2010)}]{Miklasz2010}%
  \BibitemOpen
  \bibfield  {author} {\bibinfo {author} {\bibfnamefont {K.~A.}\ \bibnamefont
  {Miklasz}}\ and\ \bibinfo {author} {\bibfnamefont {M.~W.}\ \bibnamefont
  {Denny}},\ }\href {\doibase 10.4319/lo.2010.55.6.2513} {\bibfield  {journal}
  {\bibinfo  {journal} {Limnology and Oceanography}\ }\textbf {\bibinfo
  {volume} {55}},\ \bibinfo {pages} {2513} (\bibinfo {year}
  {2010})}\BibitemShut {NoStop}%
\bibitem [{\citenamefont {Moore}\ and\ \citenamefont
  {Villareal}(1996)}]{Moore1996}%
  \BibitemOpen
  \bibfield  {author} {\bibinfo {author} {\bibfnamefont {J.~K.}\ \bibnamefont
  {Moore}}\ and\ \bibinfo {author} {\bibfnamefont {T.~A.}\ \bibnamefont
  {Villareal}},\ }\href {\doibase 10.4319/lo.1996.41.7.1514} {\bibfield
  {journal} {\bibinfo  {journal} {Limnology and Oceanography}\ }\textbf
  {\bibinfo {volume} {41}},\ \bibinfo {pages} {1514} (\bibinfo {year}
  {1996})}\BibitemShut {NoStop}%
\bibitem [{\citenamefont {Villareal}(1992)}]{Villareal1992}%
  \BibitemOpen
  \bibfield  {author} {\bibinfo {author} {\bibfnamefont {T.~A.}\ \bibnamefont
  {Villareal}},\ }\href {\doibase 10.1093/plankt/14.3.459} {\bibfield
  {journal} {\bibinfo  {journal} {Journal of Plankton Research}\ }\textbf
  {\bibinfo {volume} {14}},\ \bibinfo {pages} {459} (\bibinfo {year}
  {1992})}\BibitemShut {NoStop}%
\bibitem [{\citenamefont {Smayda}(1974)}]{Smayda1974}%
  \BibitemOpen
  \bibfield  {author} {\bibinfo {author} {\bibfnamefont {T.~J.}\ \bibnamefont
  {Smayda}},\ }\href {\doibase 10.4319/lo.1974.19.4.0628} {\bibfield  {journal}
  {\bibinfo  {journal} {Limnology and Oceanography}\ }\textbf {\bibinfo
  {volume} {19}},\ \bibinfo {pages} {628} (\bibinfo {year} {1974})}\BibitemShut
  {NoStop}%
\bibitem [{\citenamefont {Waite}\ \emph {et~al.}(1997)\citenamefont {Waite},
  \citenamefont {Fisher}, \citenamefont {Thompson},\ and\ \citenamefont
  {Harrison}}]{Waite1997}%
  \BibitemOpen
  \bibfield  {author} {\bibinfo {author} {\bibfnamefont {A.}~\bibnamefont
  {Waite}}, \bibinfo {author} {\bibfnamefont {A.}~\bibnamefont {Fisher}},
  \bibinfo {author} {\bibfnamefont {P.~A.}\ \bibnamefont {Thompson}}, \ and\
  \bibinfo {author} {\bibfnamefont {P.~J.}\ \bibnamefont {Harrison}},\
  }\href@noop {} {\bibfield  {journal} {\bibinfo  {journal} {Marine Ecology
  Progress Series}\ }\textbf {\bibinfo {volume} {157}},\ \bibinfo {pages} {97}
  (\bibinfo {year} {1997})}\BibitemShut {NoStop}%
\bibitem [{\citenamefont {Bec}\ \emph {et~al.}(2005)\citenamefont {Bec},
  \citenamefont {Celani}, \citenamefont {Cencini},\ and\ \citenamefont
  {Musacchio}}]{bec2005clustering}%
  \BibitemOpen
  \bibfield  {author} {\bibinfo {author} {\bibfnamefont {J.}~\bibnamefont
  {Bec}}, \bibinfo {author} {\bibfnamefont {A.}~\bibnamefont {Celani}},
  \bibinfo {author} {\bibfnamefont {M.}~\bibnamefont {Cencini}}, \ and\
  \bibinfo {author} {\bibfnamefont {S.}~\bibnamefont {Musacchio}},\ }\href@noop
  {} {\bibfield  {journal} {\bibinfo  {journal} {Phys. Fluids}\ }\textbf
  {\bibinfo {volume} {17}},\ \bibinfo {pages} {073301} (\bibinfo {year}
  {2005})}\BibitemShut {NoStop}%
\bibitem [{\citenamefont {Falkovich}\ \emph {et~al.}(2002)\citenamefont
  {Falkovich}, \citenamefont {Fouxon},\ and\ \citenamefont
  {Stepanov}}]{falkovich2002acceleration}%
  \BibitemOpen
  \bibfield  {author} {\bibinfo {author} {\bibfnamefont {G.}~\bibnamefont
  {Falkovich}}, \bibinfo {author} {\bibfnamefont {A.}~\bibnamefont {Fouxon}}, \
  and\ \bibinfo {author} {\bibfnamefont {M.~G.}\ \bibnamefont {Stepanov}},\
  }\href@noop {} {\bibfield  {journal} {\bibinfo  {journal} {Nature}\ }\textbf
  {\bibinfo {volume} {419}},\ \bibinfo {pages} {151} (\bibinfo {year}
  {2002})}\BibitemShut {NoStop}%
\bibitem [{\citenamefont {Squires}\ and\ \citenamefont
  {Eaton}(1991)}]{Squires1991preferential}%
  \BibitemOpen
  \bibfield  {author} {\bibinfo {author} {\bibfnamefont {K.~D.}\ \bibnamefont
  {Squires}}\ and\ \bibinfo {author} {\bibfnamefont {J.~K.}\ \bibnamefont
  {Eaton}},\ }\href@noop {} {\bibfield  {journal} {\bibinfo  {journal} {Physics
  of Fluids A: Fluid Dynamics}\ }\textbf {\bibinfo {volume} {3}},\ \bibinfo
  {pages} {1169} (\bibinfo {year} {1991})}\BibitemShut {NoStop}%
\bibitem [{\citenamefont {Bec}(2005)}]{Bec2005multifractal}%
  \BibitemOpen
  \bibfield  {author} {\bibinfo {author} {\bibfnamefont {J.}~\bibnamefont
  {Bec}},\ }\href@noop {} {\bibfield  {journal} {\bibinfo  {journal} {Journal
  of Fluid Mechanics}\ }\textbf {\bibinfo {volume} {528}},\ \bibinfo {pages}
  {255} (\bibinfo {year} {2005})}\BibitemShut {NoStop}%
\bibitem [{\citenamefont {Cencini}\ \emph {et~al.}(2006)\citenamefont
  {Cencini}, \citenamefont {Bec}, \citenamefont {Biferale}, \citenamefont
  {Boffetta}, \citenamefont {Celani}, \citenamefont {Lanotte}, \citenamefont
  {Musacchio},\ and\ \citenamefont {Toschi}}]{Cencini2006dynamics}%
  \BibitemOpen
  \bibfield  {author} {\bibinfo {author} {\bibfnamefont {M.}~\bibnamefont
  {Cencini}}, \bibinfo {author} {\bibfnamefont {J.}~\bibnamefont {Bec}},
  \bibinfo {author} {\bibfnamefont {L.}~\bibnamefont {Biferale}}, \bibinfo
  {author} {\bibfnamefont {G.}~\bibnamefont {Boffetta}}, \bibinfo {author}
  {\bibfnamefont {A.}~\bibnamefont {Celani}}, \bibinfo {author} {\bibfnamefont
  {A.~S.}\ \bibnamefont {Lanotte}}, \bibinfo {author} {\bibfnamefont
  {S.}~\bibnamefont {Musacchio}}, \ and\ \bibinfo {author} {\bibfnamefont
  {F.}~\bibnamefont {Toschi}},\ }\href {\doibase 10.1080/14685240600675727}
  {\bibfield  {journal} {\bibinfo  {journal} {Journal of Turbulence}\ }\textbf
  {\bibinfo {volume} {7}},\ \bibinfo {pages} {N36} (\bibinfo {year}
  {2006})}\BibitemShut {NoStop}%
\bibitem [{\citenamefont {Gustavsson}\ \emph {et~al.}(2016)\citenamefont
  {Gustavsson}, \citenamefont {Berglund}, \citenamefont {Jonsson},\ and\
  \citenamefont {Mehlig}}]{Gustavsson2016preferential}%
  \BibitemOpen
  \bibfield  {author} {\bibinfo {author} {\bibfnamefont {K.}~\bibnamefont
  {Gustavsson}}, \bibinfo {author} {\bibfnamefont {F.}~\bibnamefont
  {Berglund}}, \bibinfo {author} {\bibfnamefont {P.~R.}\ \bibnamefont
  {Jonsson}}, \ and\ \bibinfo {author} {\bibfnamefont {B.}~\bibnamefont
  {Mehlig}},\ }\href@noop {} {\bibfield  {journal} {\bibinfo  {journal}
  {Physical review letters}\ }\textbf {\bibinfo {volume} {116}},\ \bibinfo
  {pages} {108104} (\bibinfo {year} {2016})}\BibitemShut {NoStop}%
\bibitem [{\citenamefont {Borgnino}\ \emph {et~al.}(2018)\citenamefont
  {Borgnino}, \citenamefont {Boffetta}, \citenamefont {De~Lillo},\ and\
  \citenamefont {Cencini}}]{Borgnino2018gyrotactic}%
  \BibitemOpen
  \bibfield  {author} {\bibinfo {author} {\bibfnamefont {M.}~\bibnamefont
  {Borgnino}}, \bibinfo {author} {\bibfnamefont {G.}~\bibnamefont {Boffetta}},
  \bibinfo {author} {\bibfnamefont {F.}~\bibnamefont {De~Lillo}}, \ and\
  \bibinfo {author} {\bibfnamefont {M.}~\bibnamefont {Cencini}},\ }\href@noop
  {} {\bibfield  {journal} {\bibinfo  {journal} {Journal of Fluid Mechanics}\
  }\textbf {\bibinfo {volume} {856}} (\bibinfo {year} {2018})}\BibitemShut
  {NoStop}%
\bibitem [{\citenamefont {Redner}(2001)}]{Redner2001}%
  \BibitemOpen
  \bibfield  {author} {\bibinfo {author} {\bibfnamefont {S.}~\bibnamefont
  {Redner}},\ }\href@noop {} {\emph {\bibinfo {title} {A guide to first-passage
  processes}}}\ (\bibinfo  {publisher} {Cambridge University Press},\ \bibinfo
  {year} {2001})\BibitemShut {NoStop}%
\bibitem [{\citenamefont {Santamaria}\ \emph {et~al.}(2014)\citenamefont
  {Santamaria}, \citenamefont {De~Lillo}, \citenamefont {Cencini},\ and\
  \citenamefont {Boffetta}}]{santamaria2014gyrotactic}%
  \BibitemOpen
  \bibfield  {author} {\bibinfo {author} {\bibfnamefont {F.}~\bibnamefont
  {Santamaria}}, \bibinfo {author} {\bibfnamefont {F.}~\bibnamefont
  {De~Lillo}}, \bibinfo {author} {\bibfnamefont {M.}~\bibnamefont {Cencini}}, \
  and\ \bibinfo {author} {\bibfnamefont {G.}~\bibnamefont {Boffetta}},\
  }\href@noop {} {\bibfield  {journal} {\bibinfo  {journal} {Physics of
  Fluids}\ }\textbf {\bibinfo {volume} {26}},\ \bibinfo {pages} {111901}
  (\bibinfo {year} {2014})}\BibitemShut {NoStop}%
\bibitem [{\citenamefont {Calzavarini}\ \emph {et~al.}(2008)\citenamefont
  {Calzavarini}, \citenamefont {Cencini}, \citenamefont {Lohse},\ and\
  \citenamefont {Toschi}}]{Calzavarini2008}%
  \BibitemOpen
  \bibfield  {author} {\bibinfo {author} {\bibfnamefont {E.}~\bibnamefont
  {Calzavarini}}, \bibinfo {author} {\bibfnamefont {M.}~\bibnamefont
  {Cencini}}, \bibinfo {author} {\bibfnamefont {D.}~\bibnamefont {Lohse}}, \
  and\ \bibinfo {author} {\bibfnamefont {F.}~\bibnamefont {Toschi}},\
  }\href@noop {} {\bibfield  {journal} {\bibinfo  {journal} {Physical review
  letters}\ }\textbf {\bibinfo {volume} {101}},\ \bibinfo {pages} {084504}
  (\bibinfo {year} {2008})}\BibitemShut {NoStop}%
\bibitem [{\citenamefont {Langmuir}(1938)}]{Langmuir1938}%
  \BibitemOpen
  \bibfield  {author} {\bibinfo {author} {\bibfnamefont {I.}~\bibnamefont
  {Langmuir}},\ }\href@noop {} {\bibfield  {journal} {\bibinfo  {journal}
  {Science}\ }\textbf {\bibinfo {volume} {87}},\ \bibinfo {pages} {119}
  (\bibinfo {year} {1938})}\BibitemShut {NoStop}%
\bibitem [{\citenamefont {Thorpe}(2004)}]{Thorpe2004}%
  \BibitemOpen
  \bibfield  {author} {\bibinfo {author} {\bibfnamefont {S.~A.}\ \bibnamefont
  {Thorpe}},\ }\href@noop {} {\bibfield  {journal} {\bibinfo  {journal} {Annu.
  Rev. Fluid Mech.}\ }\textbf {\bibinfo {volume} {36}},\ \bibinfo {pages} {55}
  (\bibinfo {year} {2004})}\BibitemShut {NoStop}%
\bibitem [{\citenamefont {Barstow}(1983)}]{Barstow1983}%
  \BibitemOpen
  \bibfield  {author} {\bibinfo {author} {\bibfnamefont {S.~F.}\ \bibnamefont
  {Barstow}},\ }\href@noop {} {\bibfield  {journal} {\bibinfo  {journal} {Mar.
  Environ. Res.}\ }\textbf {\bibinfo {volume} {9}},\ \bibinfo {pages} {211}
  (\bibinfo {year} {1983})}\BibitemShut {NoStop}%
\bibitem [{\citenamefont {Lindemann}\ \emph {et~al.}(2017)\citenamefont
  {Lindemann}, \citenamefont {Visser},\ and\ \citenamefont
  {Mariani}}]{Lindemann2017}%
  \BibitemOpen
  \bibfield  {author} {\bibinfo {author} {\bibfnamefont {C.}~\bibnamefont
  {Lindemann}}, \bibinfo {author} {\bibfnamefont {A.}~\bibnamefont {Visser}}, \
  and\ \bibinfo {author} {\bibfnamefont {P.}~\bibnamefont {Mariani}},\
  }\href@noop {} {\bibfield  {journal} {\bibinfo  {journal} {J. Royal Soc.
  Interface}\ }\textbf {\bibinfo {volume} {14}},\ \bibinfo {pages} {20170453}
  (\bibinfo {year} {2017})}\BibitemShut {NoStop}%
\end{thebibliography}
%
	
\end{document}